\documentclass[aps,pra,onecolumn,10pt,superscriptaddress,nofootinbib]{revtex4-2}
\usepackage{amsmath,amssymb,bm}
\usepackage{graphicx}
\usepackage{booktabs,array}
\usepackage{tabularray}
\usepackage{tabularx}
\UseTblrLibrary{booktabs}
\usepackage{xcolor}
\usepackage{hyperref}
\usepackage{microtype}
\usepackage{flafter}
\usepackage{enumitem}
\usepackage{siunitx}
\newcommand{\uas}{\,\mu\mathrm{as}}
\newcommand{\Efin}{\mathcal{E}_{\rm fin}}

\newenvironment{measurementstatement}{%
  \begin{quote}\small\setlength{\parskip}{2pt}\setlength{\parindent}{0pt}}
 {\end{quote}}
\newcommand{\msitem}[2]{\emph{#1:} #2\par}
\newcommand{\EarthAcceptedGain}{1.033}
\newcommand{\EarthAcceptedGainLo}{1.031}
\newcommand{\EarthAcceptedGainHi}{1.045}
\newcommand{\EarthFixedGain}{1.010}
\newcommand{\EarthFixedGainLo}{1.005}
\newcommand{\EarthFixedGainHi}{1.045}
\newcommand{\EarthPassChange}{1.9}
\newcommand{\EarthPassLo}{0.0}
\newcommand{\EarthPassHi}{4.4}
\newcommand{\EarthClassMass}{10.66}
\newcommand{\EarthSqMass}{10.55}
\newcommand{\EarthClassPass}{41.2}
\newcommand{\EarthSqPass}{43.1}
\newcommand{\EarthSensitivityMedian}{1.008}
\newcommand{\EarthSensitivityPOneSix}{0.991}
\newcommand{\EarthSensitivityPEightFour}{1.026}
\newcommand{\EarthSensitivityPTwoPointFive}{0.975}
\newcommand{\EarthSensitivityPNinetySevenPointFive}{1.043}
\newcommand{\EarthProbBreakEven}{68.0}
\newcommand{\EarthProbTwo}{25.8}
\newcommand{\EarthProbFive}{1.0}
\newcommand{\QHetSdGain}{1.50}
\newcommand{\QHetFIGain}{2.24}
\newcommand{\NullerPppm}{10.7}
\newcommand{\NoonMaxGain}{0.253}
\newcommand{\CoronagraphSpeed}{2.06}
\newcommand{\CoronagraphPrecision}{1.43}
\newcommand{\QHetDirectFI}{1.41\times10^{-5}}
\newcommand{\NetworkUncYield}{3.864\times10^{-17}}
\newcommand{\SpadeGain}{1.77}
\newcommand{\SpadeGainLo}{1.47}
\newcommand{\SpadeGainHi}{2.15}

\newcommand{\SpadeTrials}{160}
\newcommand{\SpadeBootstrapSeed}{2035}
\newcommand{\SpadeDirectRMSE}{32.1}
\newcommand{\SpadeRMSE}{18.1}
\newcommand{\SpadeDirectCatCount}{17}
\newcommand{\SpadeCatCount}{3}
\newcommand{\SpadeDriftGain}{1.20}
\newcommand{\SpadeDriftGainLo}{0.89}
\newcommand{\SpadeDriftGainHi}{1.61}
\newcommand{\SpadeDriftCoverage}{41.7}
\newcommand{\SpadeDriftCoverageLo}{28.8}
\newcommand{\SpadeDriftCoverageHi}{55.7}
\newcommand{\SpadeDriftCoveredCount}{20}
\newcommand{\SpadeMedAEGain}{1.43}
\newcommand{\SpadeMedAEGainLo}{1.15}
\newcommand{\SpadeMedAEGainHi}{1.87}
\newcommand{\NetworkMeanAbsV}{6.17e-05}
\newcommand{\NetworkRmsV}{6.88e-05}
\newcommand{\NetworkFirstNull}{8.81}
\newcommand{\ValidatedVdetShort}{0.402}
\newcommand{\ValidatedQuadratureGainShort}{1.58}
\newcommand{\ValidatedReadoutBetaShort}{0.510}
\newcommand{\ValidatedReadoutGainShort}{1.20}

\hypersetup{%
  hidelinks,%
  pdftitle={Do Quantum Measurement Advantages Survive to Astrophysical Inference? Seven Benchmarks in Optical Interferometry and Imaging},%
  pdfauthor={Slava G. Turyshev},%
  pdfsubject={Astrophysical opportunities for quantum technologies in optical interferometry and high-angular-resolution imaging},%
  pdfkeywords={optical interferometry, high-angular-resolution imaging, quantum optics, exoplanet astrometry, spatial-mode demultiplexing, stellar dynamical masses}%
}

\hypersetup{colorlinks=true,allcolors=blue}

\begin{document}
\graphicspath{{figures/}}
\raggedbottom

\title{Do Quantum Measurement Advantages Survive to Astrophysical Inference? \\ Seven Benchmarks in Optical Interferometry and Imaging}
\author{Slava G. Turyshev}
%\email{slava.g.turyshev@jpl.nasa.gov}
\affiliation{Jet Propulsion Laboratory, California Institute of Technology, 4800 Oak Grove Drive, Pasadena, California 91109-0899, USA}
\date{\today}

\begin{abstract}
Quantum techniques matter to optical astronomy when they preserve information in the parameter-bearing measurement channel after realistic loss, calibration, covariance, and inference are propagated to the final astrophysical observable. We compare seven approaches by the quantity they change: internal optical-path delay, sub-Rayleigh source separation, coronagraphic planet throughput and stellar leakage, coherent field quadratures, nonlocal complex visibility, null depth, and active phase sensing. In an Earth-analog astrometry case, 6-dB squeezed internal metrology improves the measured phase quadrature by 1.58 and the complete readout by about 1.20, but fixed-time planet-mass precision by only \EarthFixedGain\ (95\% interval \EarthFixedGainLo--\EarthFixedGainHi); mission duration, useful baseline, cadence, and calibration have greater modeled leverage. For a close young binary, signed spatial-mode demultiplexing preserves sub-Rayleigh orbital information that is substantially degraded in the tested direct-imaging comparator. A representative matched-model calculation gives a dynamical-mass RMSE gain \SpadeGain\ (\SpadeGainLo--\SpadeGainHi). For illustrative equal-prior pre-main-sequence predictions at 0.10 and \(0.12M_\odot\), mass precisions of 15\%, 10\%, and 7\% correspond to correct-selection probabilities of $\sim$73\%, 82\%, and 90\%. The numerical gain remains provisional because the event-level continuous-image comparator, strict tail convergence, externally measured broadband transfer matrix, and finite-photon calibration are not yet closed. Modal coronagraphy and quantum-enhanced heterodyne reception define conditional receiver opportunities, while the nonlocal, nuller, and prepared-NOON cases identify physical bottlenecks that prevent an astrophysical advantage in the stated configurations. The common result is that quantum technology is most valuable when it changes an information-bearing observable and that channel remains consequential in the final science measurement.
\end{abstract}

\maketitle

\section{Astrophysical observables and quantum opportunities}
\label{sec:intro}
Optical interferometry and high-angular-resolution imaging infer physical source parameters through several distinct observables. A Michelson or Fizeau array estimates the complex visibility \(\gamma=V e^{i\phi}\): visibility amplitude constrains angular size and brightness structure, phase and differential phase encode photocenter shifts and asymmetric line emission, and closure quantities support synthesis imaging. A nulling interferometer suppresses the on-axis stellar field and measures residual leakage as a function of wavelength and rotation. Coherent receivers measure field quadratures, while a filled aperture records photon positions or spatial-mode occupancies. Internal metrology measures optical path delay (OPD), phase reference, pointing, and frequency transfer rather than astronomical photons directly~\cite{Monnier2003,Rajagopal2024}.

The relevant limitation is therefore observable dependent. Narrow-angle astrometry can be limited by OPD readout, baseline-vector knowledge, chromatic phase, reference geometry, and stellar photocenter motion. Close-binary imaging loses separation information near the Rayleigh limit even with fine pixels. High-contrast imaging trades accepted planet photons against finite-star, instrumental, and modal leakage. Mid-infrared nulling depends on residual piston and astrophysical backgrounds. Long-baseline coherent or nonlocal arrays depend on source visibility, phase referencing, and field or quantum-state transport. A credible quantum claim must be stated in the units of the source parameter or observing time and compared with the strongest classical measurement that uses the same astronomical resource.

Three categories of technology enter this discussion. Nonclassical resources such as squeezed vacuum, entanglement, memories, and prepared non-Gaussian states modify the state or supply ancillary correlations. Quantum-estimation-informed measurement bases such as spatial-mode demultiplexing (SPADE), SLIVER~\cite{NairTsang2016SLIVER}, and modal coronagraphy reorganize ordinary astronomical light into outputs with a more favorable information distribution~\cite{TsangNairLu2016,Tsang2019Moments,Lupo2020}. Quantum-enabled hardware---including SNSPDs, TESs, MKIDs, frequency conversion, combs, and optical clocks---can improve efficiency, timing, spectral access, or stability without by itself creating a nonclassical information advantage~\cite{Lita2008TES,Mazin2013ARCONS,Walter2020MEC,LiSNSPD2025,Caldwell2023TimeTransfer}.

Squeezed interferometry follows the unused-port noise mechanism identified by Caves~\cite{Caves1981}. Optical loss mixes ordinary vacuum back into the receiver, and general noisy-quantum-metrology bounds show why an ideal scaling need not survive realistic loss~\cite{Escher2011,Demkowicz2012}. In the astrometric benchmark below, the astronomical light is unchanged: squeezing acts only on internal differential-delay metrology, and the science question is whether that reduction changes a planet-mass posterior after the full covariance and observing program are included.

SPADE addresses a different physical loss of information. For two incoherent sources well inside the diffraction width, the direct-image intensity changes only weakly with separation, whereas point-spread-function-matched modal occupancies remain separation sensitive. The potential astronomy product is not a receiver error bar but a signed two-dimensional orbit, dynamical mass, and discrimination among pre-main-sequence stellar models. Recent on-sky binary-source hypothesis testing with binary SPADE~\cite{Wallis2026OnSkySPADE} and adaptive quantum-optimal measurements for high-contrast scenes~\cite{Choi2026AdaptiveExoplanet} establish closely related experimental directions. The present calculation is distinct: it propagates a fixed broadband signed registration-plus-modal receiver through multi-epoch Keplerian inference, transfer perturbations, and an explicit direct-imaging comparator.

The two deepest benchmarks therefore connect receiver physics to astrophysical masses. Squeezed internal metrology is tested against an Earth-analog planet mass, and signed SPADE is tested against a young-binary dynamical mass. The remaining cases define opportunities and boundaries. Modal coronagraphy can improve the planet-photon throughput--leakage balance near the inner working angle. Quantum-enhanced heterodyne reception can improve coherent-amplitude precision where heterodyne is selected for spectral or architectural reasons. Entanglement-assisted reception could avoid direct transport of the astronomical field, but only if source coherence, phase diversity, false-event control, pair delivery, memory operation, and completed yield close simultaneously~\cite{Tsang2011,Gottesman2012,Khabiboulline2019,Brown2023,Stas2026,WangMemory2026}. A squeezed null sensor and a canonical NOON probe are retained as controls that show how a receiver gain can fail to reach an astrophysical observable.

The paper asks four astronomy-facing questions. Which observable does each technique improve? Which science measurement could benefit? Under what physical and instrumental conditions does the advantage survive? What experiment would establish the next astrophysical evidence level? Sections~\ref{sec:earth} and \ref{sec:spade} treat Earth-analog and young-binary masses. Section~\ref{sec:maps} covers receiver-level opportunities for exoplanet imaging and coherent reception. Section~\ref{sec:network} evaluates the present nonlocal architecture against source coherence and visibility inference, and Sec.~\ref{sec:controls} gives negative controls. Detailed likelihood definitions, comparator grids, convergence records, parameter provenance, and resampling specifications are provided in Appendices~\ref{app:comparators}--\ref{app:controls} and in the accompanying reproducibility archive.
\section{Astrophysical opportunity map and survival framework}
\label{sec:framework}

\subsection{Which observables can quantum techniques improve?}
Table~\ref{tab:orientation} summarizes the astronomy-facing role of each technique. The rows are deliberately organized by observable and science objective rather than by statistical method.

\small
\setlength{\tabcolsep}{3pt}\renewcommand{\arraystretch}{1.08}
\begin{longtblr}[caption={Astrophysical opportunity map. ``Present status'' is the highest level reached by the calculation in this paper, not a universal maturity ranking.},label={tab:orientation}]{width=\textwidth,colspec={X[13,l] X[19,l] X[27,l] X[33.5,l] X[16.5,l]},rowhead=1}
\toprule
Technique & Observable improved & Astrophysical objective & Conditions for the advantage to survive & Present status \\\midrule
Squeezed internal metrology & Differential OPD and phase & Reflex astrometry and planet-mass characterization & High optical efficiency, stable squeeze angle, strong coupling to the final readout, low zero-point/ activity covariance, and adequate duty & End-to-end Earth-analog mass benchmark; small fixed-time gain \\
Signed SPADE & Two-dimensional separation from modal occupancies plus registration & Orbits and dynamical masses of close binaries below the ordinary imaging scale & Throughput, registration, transfer calibration, crosstalk, component labeling, chromatic stability, and a converged direct-image comparator & Matched-model young-binary mass benchmark \\
Modal coronagraphy & Accepted planet and stellar photon rates & Direct imaging or spectroscopy near the inner working angle & Planet throughput must compensate finite-star, wavefront, chromatic, and modal leakage relative to a conventional frontier & Receiver-level conditional opportunity \\
Quantum-enhanced heterodyne & Complex amplitude and spectrum & Coherent spectro-interferometry where heterodyne is selected for resolution or architecture & Squeezing must survive loss and reference distribution; source occupation must justify coherent reception relative to direct detection & Receiver-level gain over classical heterodyne only \\
Entanglement-assisted reception & Complex visibility without transporting the astronomical photon to a common combiner & Longer-baseline angular structure and source parameters & Phase-diverse visibility likelihood, useful correlated flux, false-event control, completed operations, pairs, memories, and phase reference & Source/protocol negative control in the present stress case \\
Squeezed null sensing & Residual OPD and null depth & Mid-infrared planet continuum or spectroscopy & Sensor noise must materially limit the closed-loop residual and the planet likelihood & Instrument-level negative control \\
Prepared NOON probe & Active phase response & No direct passive-astronomy science product in the present benchmark & Preparation, heralding, all-photon survival, phase diffusion, duty, and attempted-photon accounting & Receiver-level negative control \\
\bottomrule
\end{longtblr}
\normalsize

\subsection{From receiver improvement to astrophysical value}
A useful organizing chain is
\begin{equation}
G_{\rm fundamental}\rightarrow G_{\rm receiver}\rightarrow G_{\rm instrument}
\rightarrow G_{\rm inference}\rightarrow\Delta U_{\rm science}.
\label{eq:Gainchain}
\end{equation}
The first step asks whether the quantum state or measurement basis contains more information. The next steps ask whether that information survives loss, calibration, controls, observing duty, nuisance parameters, and the target likelihood. The final step asks whether the surviving precision changes an astrophysical measurement, classification, accessible angular scale, exposure time, or target set. A technique can therefore be physically successful at the receiver and still be scientifically unimportant if another term controls the final observable \cite{Turyshev_QTech2026}.

Throughout the paper, the headline quantity is the astrophysical or observing consequence: planet-mass precision, binary dynamical-mass recovery, high-contrast exposure time, complex-amplitude information, visibility sensitivity, or nuller planet-amplitude precision. The common likelihood notation, resource denominators, sensitivity diagnostics, convergence record, and uncertainty definitions are collected in Appendices~\ref{app:comparators}--\ref{app:spade}.

\subsection{Compact notation and terminology}
Table~\ref{tab:notation} defines the abbreviations and symbols used across the otherwise distinct receiver architectures.

\small
\setlength{\tabcolsep}{5pt}\renewcommand{\arraystretch}{1.12}
\begin{longtblr}[caption={Notation and terminology used throughout.},label={tab:notation}]{width=\textwidth,colspec={X[16,l] X[28,l] X[48,l]},rowhead=1}
\toprule
Term or symbol & Meaning & Role in the analysis \\\midrule
Quadrature & One of two conjugate field-amplitude coordinates & Homodyne phase readout; squeezing lowers one variance and raises the other \\
\(V_{\rm det}\) & Detected squeezed-quadrature variance normalized to vacuum & Receiver-level variance after loss and angle jitter \\
SPADE & Spatial-mode demultiplexing & Sorts incoherent photons into PSF-matched modes \\
HG & Hermite--Gaussian & Modal basis used in the Gaussian-PSF SPADE benchmark \\
PSF & Point-spread function & Image and modal response of the telescope \\
RMSE & Root-mean-square error & Tail-sensitive empirical estimator metric \\
MedAE & Median absolute error & Robust central empirical estimator metric \\
ASD & Amplitude spectral density & Square root of one-sided power spectral density \\
HBT & Hanbury Brown--Twiss & Intensity-correlation interferometry \\
SNSPD/TES/MKID & Superconducting detector technologies & Quantum-enabled hardware context, not automatically a nonclassical advantage \\
SLIVER & Superlocalization by image inversion interferometry & Quantum-informed image measurement basis \\
Heralding & Classical indication that a quantum operation/event succeeded & Converts some faults into erasures but consumes time and resources \\
Quantum memory & Device storing a quantum state until a later operation & Required in memory-assisted nonlocal reception \\
NOON state & \((|N,0\rangle+|0,N\rangle)/\sqrt2\) & Canonical prepared path-entangled phase probe \\
\bottomrule
\end{longtblr}
\normalsize

\section{Squeezed internal metrology for Earth-analog mass measurements}
\label{sec:earth}

The science question is whether quantum-limited internal metrology can materially improve the mass of a nearby Earth analog once stellar activity, reference-star geometry, orbital information, calibration, and observing duty are included. The result is intentionally expressed in planet-mass precision rather than in detected squeezing alone.

\begin{measurementstatement}
\msitem{Astrophysical observable}{Differential delay and angular reflex motion of a solar analog at 10 pc.}
\msitem{Limiting classical effect}{A quantum-coupled readout component embedded in a larger covariance containing zero point, reference geometry, baseline/attitude, stellar photocenter, thermal/chromatic, and orbital-prior terms.}
\msitem{Quantum intervention}{Six-decibel phase-squeezed vacuum in the unused port of the internal differential-delay metrology interferometer; astronomical starlight is unchanged.}
\msitem{Science-facing metric}{Detected quadrature variance, complete-readout variance, and paired planet-mass posterior precision.}
\msitem{Required classical comparator}{An optimized unsqueezed homodyne receiver with the same carrier, band, detector chain, accepted-time likelihood, and fixed elapsed time, plus classical mission and calibration interventions evaluated in the same final mass metric.}
\msitem{Conditions for astrophysical gain}{High optical efficiency, stable squeeze angle, a large physically coupled readout fraction, adequate zero-point coupling, low activity covariance, and squeezed-system duty close to the classical duty.}
\msitem{Present scientific status}{Inference-level paired nonlinear simulation; the conditional mass-requirement change is reported as a decision proxy with paired uncertainty.}
\end{measurementstatement}

This benchmark is a conditional mass-characterization experiment for a known target and modeled one-year orbit.  It does not evaluate blind detection completeness, blind orbit recovery, or mission yield for previously unknown Earth twins.  The orbit period, phase, and orientation information supplied by the declared RV and direct-imaging precursor program is part of the resource accounting; weakening those priors reduces characterization performance for both receivers and can dominate the squeezing response.

\subsection{Physical mechanism and receiver model}
A homodyne phase receiver measures one field quadrature relative to a bright coherent carrier.  Ordinary vacuum fluctuations entering the unused port contribute to the phase-noise variance.  A squeezed vacuum state redistributes uncertainty between conjugate quadratures: the measured phase quadrature is reduced, while the amplitude quadrature is antisqueezed.  Optical loss is equivalent to mixing in ordinary vacuum, and squeeze-angle jitter projects some antisqueezed noise onto the measured phase quadrature.  The present receiver is an active internal metrology interferometer spanning a modeled \(1\,\mathrm{mHz}\)--\(100\,\mathrm{Hz}\) band; it measures differential delay and supports fringe tracking and astrometric calibration.  It does not squeeze or otherwise modify the passive astronomical field.

For squeeze parameter \(r\), normalized quadrature variances \(V_-=e^{-2r}\), \(V_+=e^{2r}\), total efficiency \(\eta\), and Gaussian angle jitter \(\sigma_\psi\),
\begin{align}
V_{\rm rot}&=\frac{V_-+V_+}{2}+\frac{V_--V_+}{2}e^{-2\sigma_\psi^2},\\
V_{\rm det}&=(1-\eta)+\eta V_{\rm rot}.
\label{eq:squeeze}
\end{align}
Six-decibel input squeezing, \(\eta=0.80\), and \(\sigma_\psi=0.02\) rad give \(V_{\rm det}=\ValidatedVdetShort\) and a phase-quadrature standard-deviation gain \(V_{\rm det}^{-1/2}=\ValidatedQuadratureGainShort\).  The complete 35-pm readout contains a 25-pm quantum-coupled component plus detector/electronic, laser-technical, calibration-transfer, backscatter, and other technical phase terms.  Under the independent-component approximation,
\begin{equation}
\beta_{\rm read}=\frac{25^2}{35^2}=\ValidatedReadoutBetaShort,
\qquad
\sigma_{\rm read,q}^2=\sigma_{\rm read}^2[1-\beta_{\rm read}(1-V_{\rm det})].
\label{eq:beta}
\end{equation}
The coefficient \(\beta_{\rm read}\) is an effective coupling coefficient; correlations or common-mode response can move it in either direction.  Only 9\% of the 40-pm epoch zero-point variance is assumed to share the enhanced optical channel.  The modeled complete-readout gain is therefore \ValidatedReadoutGainShort\ before the remaining astrometric covariance is introduced.

The valid-operation model contains phase-reference acquisition, calibration availability, reset/recovery, squeezed-source lock, and squeeze-angle lock.  Multiplying the declared factors gives \(d_{\rm cl}=0.980\) and \(d_q=0.932\).  Fringe-order rejection is applied separately and identically.  Accepted-data comparisons condition on valid observations; fixed-time comparisons apply architecture-specific masks to the common candidate schedule.

\subsection{Source, observing program, and likelihood}
The science target is a \(1M_\oplus\), 1-AU planet around a \(1M_\odot\), \(1R_\odot\), 5772-K star at 10 pc.  For \(M_p\ll M_\star\) and \(e=0\),
\begin{equation}
\alpha_\star=\frac{M_p}{M_\star}\frac{a_p}{d}=0.300\uas,
\label{eq:reflex}
\end{equation}
which corresponds to 146 pm on a 100-m projected baseline.  The wavelength-dependent stellar flux uses the physical radius--distance normalization
\begin{equation}
F_\nu(\lambda)=\pi B_\nu(5772\,\mathrm{K})
\left(\frac{R_\odot}{10\,\mathrm{pc}}\right)^2.
\label{eq:earthsed}
\end{equation}
At 0.85, 0.95, 1.05, and \(1.15\,\mu\mathrm m\), the flux densities are 58.12, 57.86, 56.27, and 53.92 Jy.  The uniform-disk source visibility,
\begin{equation}
V_{\rm src}(B,\lambda)=\frac{2J_1(\pi B\theta_\star/\lambda)}{\pi B\theta_\star/\lambda},
\qquad V_{\rm obs}=v_{\rm inst}V_{\rm src},
\label{eq:earthvis}
\end{equation}
uses \(\theta_\star=0.93\) mas and a separate instrumental contrast \(v_{\rm inst}=0.90\).  At \(B=100\) m, \(V_{\rm src}=0.691,0.747,0.789,0.822\).  One 1-m collector, 20\% throughput, a 10-nm channel, and a 900-s accepted epoch give detected rates per telescope of \(1.62,1.44,1.27,1.11\times10^6\,\mathrm{s^{-1}}\).  The target and three references with AB magnitudes 8.5, 9.3, and 10.0 contribute separately to the differential phase covariance.

The five-year program has 60 candidate epochs and two orthogonal projected-baseline coordinates.  The model contains position, proper motion, parallax, a projected Keplerian reflex orbit, reference-star parameters, baseline scale and orientation, attitude, chromatic delay, readout, epoch zero point, thermal drift, and a two-dimensional stellar-photocenter process.  The nominal photocenter kernel is
\begin{equation}
K_{\rm pc}(t,t')=\sigma_{\rm pc}^2 e^{-|t-t'|/\tau_{\rm pc}},
\qquad
\sigma_{\rm pc}=0.05\uas,
\quad \tau_{\rm pc}=25\,\mathrm d.
\label{eq:pckernel}
\end{equation}
The adopted activity scale is a controlled benchmark informed by solar and starspot astrometric-jitter studies rather than a universal quiet-Sun floor~\cite{Makarov2009Starspot,Deagan2026SolarJitter}.  The 20-pm thermal component has a 60-d correlation time; baseline/reference, attitude, and frame floors are 0.035, 0.025, and \(0.025\uas\) equivalent.  A 15-pm chromatic term and 5-pm detector floor are retained.

External orbital information is an observational resource rather than a free prior.  The primary radial-velocity (RV) campaign contains 120 visits over five years, a 0.5-m s\(^{-1}\) independent floor, 1.6-m s\(^{-1}\) correlated activity amplitude, 25-d activity coherence, 25.4-d rotation, and 62\% seasonal visibility, giving an effective \(\sigma_K\simeq0.10\) m s\(^{-1}\).  Direct imaging supplies 5-deg priors on inclination and node, a 0.03-rad phase prior, and \(\sigma_P/P=0.001\).  The low-mass RV model is
\begin{equation}
K\simeq\left(\frac{2\pi G}{P}\right)^{1/3}
\frac{M_p\sin i}{M_\star^{2/3}\sqrt{1-e^2}},
\label{eq:rv}
\end{equation}
valid for \(M_p\ll M_\star\); the injected orbit is circular.  The nonlinear fit profiles linear astrometric and reference parameters and optimizes amplitude, inclination, node, phase, and period.  Near-term and optimistic campaigns with effective \(\sigma_K\simeq0.28\) and 0.03 m s\(^{-1}\) are retained as sensitivity cases. Tables~\ref{tab:earthprecursor} and \ref{tab:earth} summarize, respectively, the precursor information and the canonical end-to-end Earth-analog benchmark.

\begin{table}[tb]
\caption{Information supplied before interferometric inference.  All mass uncertainties are medians from the 160-realization ensembles; ``precursor only'' uses the RV and orbit priors without interferometric astrometry.  The orientation, phase, and period constraints are illustrative direct-imaging/precursor-astrometry inputs shared by the three RV scenarios.}
\label{tab:earthprecursor}
\small
\setlength{\tabcolsep}{4pt}\renewcommand{\arraystretch}{1.12}
\begin{tabular}{@{}lrrrrr@{}}
\toprule
RV program & Total RV visits / campaign span (yr) & \(\sigma_K\) (m s\(^{-1}\)) & Precursor only & Classical & Squeezed \\
\midrule
Near term & 80 / 5 & 0.28 & 378\% & 10.56\% & 10.39\% \\
Nominal next generation & 120 / 5 & 0.10 & 126\% & 10.66\% & 10.55\% \\
Optimistic & 300 / 5 & 0.03 & 37.3\% & 10.14\% & 9.94\% \\
\bottomrule
\end{tabular}
\par\smallskip
All cases also use 5-deg inclination and node priors, a 0.03-rad phase prior, and \(\sigma_P/P=0.001\).  The interferometer supplies most of the mass information; squeezing supplies the final percent-level increment.
\end{table}

\begin{table}[tb]
\caption{Canonical Earth-twin benchmark.  The primary paired ensemble contains four seeds and 160 paired realizations per architecture.  Angular-equivalent terms are projected into delay with the baseline geometry.}
\label{tab:earth}
\small
\setlength{\tabcolsep}{4pt}\renewcommand{\arraystretch}{1.13}
\begin{tblr}{width=\textwidth,colspec={X[16.5,l] X[36,l] X[38,l]}}
\toprule
Group & Canonical values & Role \\
\midrule
Source and planet & \(T_{\rm eff}=5772\) K, \(R_\star=M_\star=1\) solar, \(d=10\) pc; \(M_p=1M_\oplus\), \(a=1\) AU, \(P=1\) yr, \(e=0\) & Reflex amplitude, physical SED, source visibility, and orbit \\
Optics & Two 100-m projected baselines; one 1-m collector per arm; four 10-nm channels at 0.85--1.15 \(\mu\)m; throughput 0.20 & Channel-resolved differential phase and photon covariance \\
Cadence & 60 candidate epochs over 5 yr; 900 s per accepted epoch & Fixed-time epoch geometry and dropout \\
Readout/calibration & 35-pm readout; 40-pm zero point; 15-pm chromatic; 20-pm thermal; 5-pm detector floor & Instrument covariance and quantum-accessible components \\
Astrophysics/geometry & \(\sigma_{\rm pc}=0.05\uas\), \(\tau_{\rm pc}=25\) d; baseline/reference 0.035, attitude 0.025, frame 0.025 \(\mu\)as & Nuisance marginalization and model-mismatch tests \\
Squeezing & 6 dB, \(\eta=0.80\), \(\sigma_\psi=0.02\) rad, \(\beta_{\rm read}=0.510\), zero-point coupling 0.09; duties 0.980 and 0.932 & Receiver variance and fixed-time schedule \\
External program & 120 RV visits over 5 yr; \(\sigma_K\simeq0.10\) m s\(^{-1}\); 5-deg orientation, 0.03-rad phase, 0.1\% period priors & Physical precursor information for characterization \\
\bottomrule\end{tblr}
\end{table}

\subsection{Astrophysical impact on planet-mass characterization}
\begin{figure}[!htbp]
\centering
\includegraphics[width=0.98\textwidth]{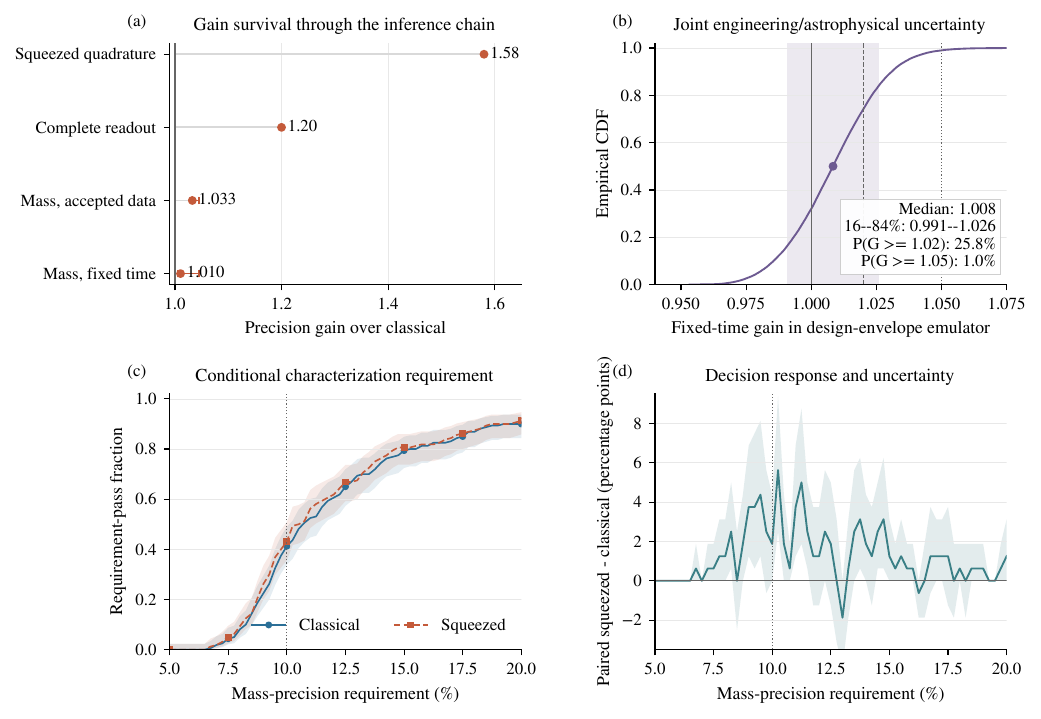}
\caption{Earth-twin gain survival and uncertainty.  (a) Receiver-to-inference attenuation: 6-dB squeezed-quadrature gain, complete-readout gain, and paired accepted-data and fixed-time planet-mass precision gains; the final two intervals are paired-bootstrap 95\% intervals.  (b) Empirical CDF of fixed-time gain from 60,000 finite-intervention design-envelope samples; the shaded band spans the 16th--84th percentiles, and the dashed line marks unity.  This is a calibrated sensitivity emulator, not a second posterior ensemble.  (c) Conditional mass-requirement pass fraction versus required mass precision; Wilson 95\% bands use the paired nonlinear ensemble.  (d) Paired squeezed-minus-classical pass-fraction change with a bootstrap 95\% band.  The figure separates a secure precision response from an unresolved thresholded decision change.}
\label{fig:earthGain}
\end{figure}

Figure~\ref{fig:earthGain}a presents the complete gain chain.  The 1.58 phase-quadrature gain becomes 1.20 for the complete readout, \(G_{M,\rm acc}=\EarthAcceptedGain\) with interval \([\EarthAcceptedGainLo,\EarthAcceptedGainHi]\), and \(G_{M,\rm fixed}=\EarthFixedGain\) with interval \([\EarthFixedGainLo,\EarthFixedGainHi]\).  Median fixed-time mass uncertainty changes from \(\EarthClassMass\%\) to \(\EarthSqMass\%\).  The conditional fraction satisfying \(\sigma_M/M<10\%\) changes from \(\EarthClassPass\%\) to \(\EarthSqPass\%\), but the paired change \(\EarthPassChange\) points has interval \([\EarthPassLo,\EarthPassHi]\).  The mass precision improves, but the change is too small to establish a different characterization outcome at the stated threshold.

With an independent radius prior, bulk density obeys
\begin{equation}
\left(\frac{\sigma_\rho}{\rho}\right)^2=
\left(\frac{\sigma_M}{M}\right)^2+9\left(\frac{\sigma_R}{R}\right)^2.
\label{eq:density}
\end{equation}
For a 2\% radius prior, median density uncertainty changes from 12.23\% to 12.14\%.  This change is too small to establish a robust composition or target-priority decision in the present benchmark.  The exact one-year blind astrometric orbit also contains a parallax-degenerate direction; squeezing reduces measurement noise but cannot restore a missing information rank.

\subsection{What controls the science return?}
The nominal \(1.010\) result is not universal.  To expose the dependence on engineering and astrophysical allocation, we draw 60,000 triangular samples over 3--10 dB squeezing, \(0.68\le\eta\le0.92\), 0.004--0.055-rad angle jitter, readout coupling 0.25--0.82, zero-point coupling 0--0.30, squeezed duty 0.86--0.985, readout and zero-point scales, photocenter scale 0.5--2.2, and thermal scale 0.5--1.6.  The model updates the ratio of quantum-coupled covariance to total covariance and is calibrated to the full nonlinear accepted-data intervention; it is an explicitly labeled design-envelope emulator, not a replacement orbit posterior.

The resulting fixed-time gain has median \(\EarthSensitivityMedian\), 16th--84th percentiles \(\EarthSensitivityPOneSix\)--\(\EarthSensitivityPEightFour\), and 2.5th--97.5th percentiles \(\EarthSensitivityPTwoPointFive\)--\(\EarthSensitivityPNinetySevenPointFive\).  The fraction above break-even is \(\EarthProbBreakEven\%\); \(\EarthProbTwo\%\) exceed 1.02, \(\EarthProbFive\%\) exceed 1.05, and none exceed 1.10.  Spearman rank correlations identify squeezed-system duty (0.80), readout coupling (0.32), photocenter scale (-0.22), zero-point coupling (0.22), and readout allocation/squeezing level (each about 0.20) as the leading controls.  These rankings describe the stated design range and should not be interpreted as independent hardware tolerances.

\begin{figure}[!htbp]
\centering
\includegraphics[width=0.98\textwidth]{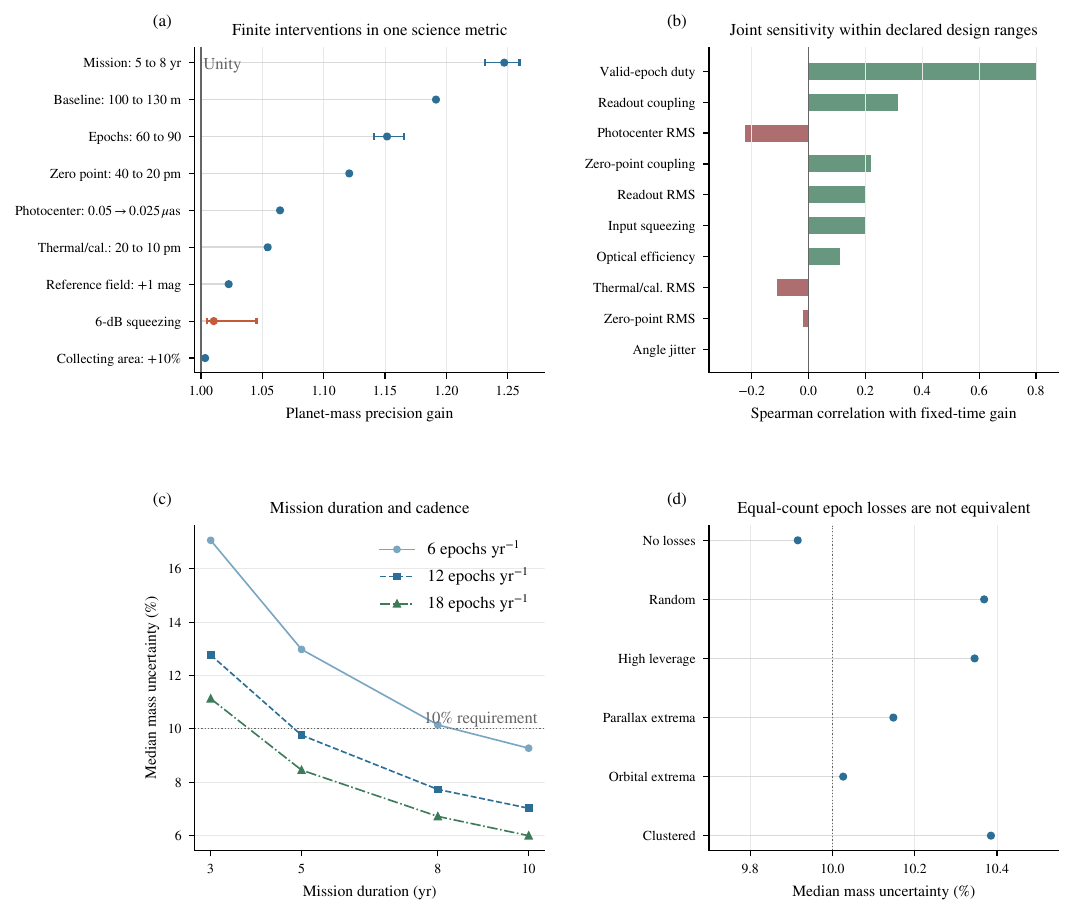}
\caption{Earth-twin science leverage and scheduling geometry.  (a) Finite intervention gains in the same planet-mass metric; intervals are shown where paired ensembles exist, and the vertical line marks unity.  (b) Spearman correlations between the declared joint-sensitivity inputs and fixed-time gain; green indicates that increasing the parameter tends to increase gain and muted red the opposite.  (c) Median mass uncertainty versus mission duration and annual cadence from deterministic expected-information reruns; the dotted line marks a 10\% requirement.  (d) Effect of removing six epochs with different temporal geometry.  The panels show that mission duration, useful baseline, cadence, and calibration have greater modeled leverage than the nominal squeezed receiver.}
\label{fig:earthdesign}
\end{figure}

Figure~\ref{fig:earthdesign}a compares finite classical and quantum interventions in the same final mass metric.  Extending the mission to eight years gives a 1.23 gain, increasing the useful baseline from 100 to 130 m gives 1.19, increasing from 60 to 90 epochs gives 1.16, halving zero-point RMS gives 1.12, halving photocenter RMS gives 1.06, halving thermal drift gives 1.05, a reference field one magnitude brighter gives 1.02, squeezing gives 1.010 at fixed time, and 10\% more collecting area gives 1.003.  These are not equal-cost actions; a common cost comparison would additionally require architecture-specific mass, power, risk, and schedule models.

A controlled epoch-mask calculation shows that removing six clustered epochs is more damaging than removing six orbital-extremum epochs.  Thus which observations are lost can matter more than the number lost.  The finite-intervention sensitivity is \(\Efin=0.104\) for accepted data but 0.034 at fixed time because duty changes the epoch mask and nuisance geometry.  A held-out doubled-photocenter ensemble reduces nominal 68\% coverage to approximately 0.63 classically and 0.61 with squeezing, a discrepancy larger than the quantum-induced decision shift.

\subsection{Interpretation}
The physical improvement is a 60\% reduction of the coupled phase-quadrature variance.  About half of the complete readout is coupled to that quadrature, reducing the full readout gain to 1.20.  Once reference-star noise, zero point, geometry, stellar activity, orbital priors, and duty enter the posterior, only a \(1.010\) fixed-time mass-precision gain remains.  The new bottleneck is therefore not the amount of squeezing but the fraction of final information that still flows through the enhanced channel.  A next evidence-level experiment must measure the end-to-end coupling coefficient, duty, zero-point sharing, and recovered astrometric parameter under realistic activity and calibration mismatch, rather than report detected squeezing alone.

\section{Signed spatial-mode detection for young-binary dynamical masses}
\label{sec:spade}

Dynamical masses of close young binaries test pre-main-sequence structure, magnetic inflation, and age scales. Ordinary images lose separation information when the components lie well inside the diffraction width, so the central science question is whether a signed modal receiver can recover a two-dimensional orbit and improve the mass precision enough to discriminate physically different stellar models.

\begin{measurementstatement}
\msitem{Astrophysical observable}{Two-dimensional relative astrometry over 24 epochs and the resulting Keplerian dynamical mass of a Taurus-like young binary.}
\msitem{Limiting classical effect}{Rayleigh information loss and multimodal tails in image-plane photon positions at separations well below the diffraction width.}
\msitem{Quantum intervention}{A quantum-estimation-informed spatial-mode basis with a signed registration image; no nonclassical illumination is used.}
\msitem{Science-facing metric}{Exact-Poisson position/modal likelihood, vector-separation error, orbital-parameter error, dynamical-mass RMSE, and illustrative pre-main-sequence model selection.}
\msitem{Required classical comparator}{Pixel-integrated direct imaging with higher detector efficiency, common latent realizations, explicit overflow counts, and controlled grid and field-of-view refinement.}
\msitem{Conditions for astrophysical gain}{Registration and transfer calibration, low nearest-neighbor crosstalk, adequate mode overlap and throughput, component labeling, stable chromatic response, and a direct-image comparator with understood nonlinear tails.}
\msitem{Present scientific status}{Inference-level paired photon-count simulation under a Gaussian PSF and simulator-internal matched calibration.}
\end{measurementstatement}

\subsection{Physical mechanism and signed receiver}
An ordinary image of two incoherent sources becomes insensitive to their separation as the sources merge: the intensity profile changes only at second order, and the direct-image separation Fisher information vanishes.  This is an information loss, not simply coarse pixel sampling.  A SPADE receiver sorts photons into point-spread-function-matched modes.  First-order modal occupancy changes remain separation sensitive in the small-separation limit, so the ideal Gaussian model has
\begin{equation}
F_Q=\frac{1}{4\sigma_{\rm PSF}^2},
\qquad
F_{\rm DI}\simeq\frac{d^2}{8\sigma_{\rm PSF}^4},
\qquad
G_{\rm ideal}=\frac{\sqrt2\sigma_{\rm PSF}}{d}.
\label{eq:spadeideal}
\end{equation}
Equation~\eqref{eq:spadeideal} is a scalar reference bound for a known centroid, known separation direction, and one separation parameter.  The implemented receiver must instead estimate a signed vector while nuisance parameters such as centroid and flux ratio are unknown.  For local parameters
\begin{equation}
\bm\theta_{\rm loc}=(d_x,d_y,c_x,c_y,f),
\end{equation}
the Poisson information matrix is
\begin{equation}
F_{ab}^{\rm rec}=\sum_j\frac{1}{\mu_j}
\frac{\partial\mu_j}{\partial\theta_a}
\frac{\partial\mu_j}{\partial\theta_b},
\qquad
F_{dd}^{\rm marg}=F_{dd}-F_{d\nu}F_{\nu\nu}^{-1}F_{\nu d},
\label{eq:spademulti}
\end{equation}
where \(d=(d_x,d_y)\), \(\nu=(c_x,c_y,f)\), and the flux-ratio prior used in the orbital recovery is included in \(F_{\nu\nu}\).  No independent centroid prior is added because the registration photons already carry that information.  At the representative first epoch and the central channel, the implemented local calculation gives the results in Table~\ref{tab:spadefim}.  The scalar ideal QFI is therefore a useful upper reference, but it is not a jointly attainable guarantee for centroid, relative brightness, both displacement components, and the Keplerian parameters.  In multiparameter quantum estimation, scalar-optimal measurements can be incompatible with simultaneous estimation of other parameters; the relevant benchmark is the full matrix and the chosen science utility, not one diagonal QFI element~\cite{TsangNairLu2016,Tsang2019Moments,Lupo2020}.

\begin{table}[tb]
\caption{Representative local multiparameter information check at the first epoch and \(1.0\,\mu\mathrm m\), with \(4.16\times10^4\) photons before detector loss and \(\sigma_{\rm PSF}=15.0\) mas.  The table reports the nuisance-marginalized two-dimensional displacement covariance.  These local values diagnose the implemented measurement basis; the headline orbital result is obtained from the nonlinear 24-epoch Poisson recovery.}
\label{tab:spadefim}
\small
\setlength{\tabcolsep}{5pt}\renewcommand{\arraystretch}{1.12}
\begin{tabular}{@{}lrrrr@{}}
\toprule
Receiver & \(\sigma(d_x)\) (mas) & \(\sigma(d_y)\) (mas) & \(\sqrt{\det C_d}\) (mas\(^2\)) & Stage \\
\midrule
Direct image, 17-by-17 & 4.93 & 6.74 & 32.9 & Local CFI \\
Signed SPADE & 0.815 & 3.83 & 3.12 & Local CFI \\
\bottomrule
\end{tabular}
\end{table}

The incoming light remains ordinary incoherent starlight.  ``Quantum-informed'' refers to the measurement basis selected from quantum-estimation theory, not to a nonclassical source.

Modal populations alone are even in displacement and do not identify a quadrant.  The modeled receiver therefore assigns 25\% of detected photons to a registration image, which identifies the brighter component and estimates the sorter origin and orientation, while 75\% enter HG\(_{00}\), HG\(_{10}\), HG\(_{01}\), and one residual channel.  At exact equal flux, component exchange is a physical symmetry; sign is then defined only modulo label exchange.  This division separates direction information from sub-Rayleigh modal information without double counting the registration photons.

For epoch \(e\), wavelength channel \(k\), registration pixel \(j\), and modal output \(m\), the joint likelihood is
\begin{equation}
\mathcal L=\prod_{e,k,j}\mathrm{Pois}(n^{\rm reg}_{ekj}|\mu^{\rm reg}_{ekj})
\prod_{e,k,m}\mathrm{Pois}(n^{\rm mod}_{ekm}|\mu^{\rm mod}_{ekm}).
\label{eq:spadelike}
\end{equation}
With relative flux \(f\), weights \(w_1=(1+f)^{-1}\), \(w_2=f(1+f)^{-1}\), signed displacement \(\bm d_e=(d_{x,e},d_{y,e})\), and calibrated origin \(\bm c_e\), the component positions are \(\bm r_{1,e}=-w_2\bm d_e-\bm c_e\) and \(\bm r_{2,e}=w_1\bm d_e-\bm c_e\).  The Poisson means are
\begin{align}
\mu^{\rm reg}_{ekj}&=N_{\gamma,k}\eta_k f_{\rm reg}
\left[w_1\Pi_j(\bm r_{1,e})+w_2\Pi_j(\bm r_{2,e})\right]+b^{\rm reg}_{ekj},\\
\bm\mu^{\rm mod}_{ek}&=N_{\gamma,k}\eta_k(1-f_{\rm reg})
\bm G_{ek}\bm R(h)\bm C_{\rm nn}(c_{ek})
\left[w_1\bm p(\bm r_{1,e})+w_2\bm p(\bm r_{2,e})\right]+\bm b^{\rm mod}_{ek}.
\label{eq:spademeans}
\end{align}
Here \(\Pi_j\) is the pixel-integrated Gaussian probability including an overflow outcome, \(\bm G_{ek}\) is a diagonal relative-gain matrix, \(\bm R(h)\) transfers unresolved retained-mode probability into the residual output, and \(\bm C_{\rm nn}\) is the physical nearest-neighbor crosstalk model.  For \(u_x=x/(2\sigma_k)\), \(u_y=y/(2\sigma_k)\), and \(q=u_x^2+u_y^2\),
\begin{equation}
\bm p=(e^{-q},u_x^2e^{-q},u_y^2e^{-q},1-e^{-q}(1+u_x^2+u_y^2))^{\mathsf T}.
\label{eq:spadeprob}
\end{equation}
The column-stochastic nearest-neighbor matrix is
\begin{equation}
\bm C_{\rm nn}(c)=
\begin{pmatrix}
1-c&c/2&c/2&0\\ c/2&1-c&0&c/2\\ c/2&0&1-c&c/2\\0&c/2&c/2&1-c
\end{pmatrix}.
\label{eq:spadeCrosstalk}
\end{equation}
It couples HG\(_{00}\) to the first-order modes and those modes to the residual output; residual leakage is parameterized independently.

\subsection{Radiometry, orbit, and calibration model}
The fiducial system is a physically defined Taurus-like young binary at 140 pc, with angular semimajor axis 1.50 mas, total mass near \(0.1M_\odot\), flux ratio 0.70 at \(1\,\mu\mathrm m\), eccentricity 0.25, inclination \(60^\circ\), node \(35^\circ\), argument of periastron \(70^\circ\), and 24 epochs over 1.2 yr.  A 6-m space telescope observes three 0.5-nm bands centered at 0.9, 1.0, and \(1.1\,\mu\mathrm m\) for 900 s per epoch with 15\% optical throughput.  A 3200-K spectrum normalized to total AB magnitude 16 gives
\begin{equation}
N_{\gamma,k}=\frac{F_{\nu,k}\Delta\nu_k A\tau_{\rm opt}t_{\rm exp}}{h\nu_k},
\label{eq:spaderad}
\end{equation}
with pre-detector counts \(3.83,4.16,4.30\times10^4\) per epoch, or \(1.23\times10^5\) total.  The Gaussian benchmark uses \(\sigma_{\rm PSF}=0.437\lambda/D\), giving 13.52, 15.02, and 16.53 mas; this matches the Airy full width at half maximum but is not a pupil-propagated modal receiver.  A local Airy/Gaussian information comparison is included in the archive, and the end-to-end claim remains limited to the Gaussian-PSF benchmark.

With 85\% SPADE detector efficiency and the 25\% registration allocation, \(N_{\rm reg}=2.61\times10^4\) detected photons per epoch and the photon-limited centroid scale is \(\sigma_{\rm PSF}/\sqrt{N_{\rm reg}}=0.0062\sigma_{\rm PSF}\).  This scale is derived from the registration likelihood and is not reintroduced as an independent prior.  The direct comparator uses 90\% efficiency and assigns all detected photons to the image likelihood.

The Kepler fit uses \((a,e,i,\Omega,\omega,M_0,P,f)\), a 0.5\% distance prior, \(\sigma_P/P=0.002\), flux-ratio prior 0.05, orientation prior 0.25 rad, and a common source/background model.  Position and modal calibration parameters are jointly represented by internally calibrated nuisance states and held-out transfer perturbations.  The orbit rotates relative to the sorter at every epoch, so the likelihood propagates signed \((d_x,d_y)\) directly rather than first forming a scalar astrometric error. Table~\ref{tab:spade} collects the canonical receiver, radiometry, transfer, prior, and ensemble assumptions used in the orbital recovery.

\begin{table}[tb]
\caption{Canonical SPADE assumptions.  The main end-to-end claim is a Gaussian-PSF, internally calibrated simulation; calibration and population mismatch are evaluated in separate finite ensembles.}
\label{tab:spade}
\small
\setlength{\tabcolsep}{4pt}\renewcommand{\arraystretch}{1.12}
\begin{tblr}{width=\textwidth,colspec={X[18,l] X[35,l] X[41,l]}}
\toprule
Quantity & Canonical value & Function \\
\midrule
Platform/cadence & 6-m space aperture; 24 epochs over 1.2 yr & Three 0.5-nm channels at 0.9--1.1 \(\mu\)m \\
Source & AB 16; 3200 K; \(a=1.50\) mas; \(e=0.25\); \(i=60^\circ\); \(f(1\,\mu\mathrm m)=0.70\) & \(1.23\times10^5\) pre-detector photons per epoch and a signed Kepler orbit \\
PSF & \(\sigma_{\rm PSF}=0.437\lambda/D\) & Gaussian benchmark matched to Airy FWHM \\
SPADE channels & 25\% registration; 75\% HG\(_{00}\), HG\(_{10}\), HG\(_{01}\), residual & Registration supplies sign/origin; modal counts supply separation information \\
Efficiencies & \(\eta_{\rm DI}=0.90\); \(\eta_{\rm SPADE}=0.85\) & Classical comparator receives the higher efficiency \\
Direct comparator & Pixel-integrated grids, overflow bin, fields to \(\pm5\sigma_{\rm PSF}\); local unbinned-information check & High-resolution nonlinear comparator and a separate local-information check \\
Static transfer & \(c=2\times10^{-4}\); residual leakage \(10^{-4}\) & Nearest-neighbor modal mixing and truncation \\
Drift & 45-d correlation; chromatic correlation 0.8 & Origin, angle, modal gain, and crosstalk processes \\
Priors & distance 0.5\%; period 0.2\%; flux ratio 0.05; orientation 0.25 rad & Identical astrophysical priors for both receivers \\
Ensembles & 160 paired primary; 48 per stress point; eight-system cluster-bootstrap population & RMSE, MedAE, coverage, catastrophic fraction, and target diversity \\
\bottomrule\end{tblr}
\end{table}

\subsection{Astrophysical result and limits of the representative benchmark}
The classical comparator was refined over 9, 11, 13, 17, and 21 pixels per axis and fields of view \(\pm3\), \(\pm4\), and \(\pm5\sigma_{\rm PSF}\), giving fifteen grid--field combinations; separate \(5\times5\) and \(7\times7\) runs test low-resolution sampling. A local \(61\times61\) calculation over \(\pm5\sigma_{\rm PSF}\) retains 99.5\% of the continuous Gaussian-image separation information, but local Fisher-information retention does not determine nonlinear orbit-recovery tails. Table~\ref{tab:spadeconv} reports the available common-realization nonlinear sequences; the complete receiver-specific convergence record is given in Table~\ref{tab:spadetail}.

\begin{table}[tb]
\caption{Paired nonlinear direct-imaging benchmarks at $\pm5\sigma_{\rm PSF}$. RMSE and MedAE are fractional dynamical-mass errors; gains use the same common realizations and brackets are paired-bootstrap 95\% percentile intervals. The receiver-specific strict convergence record is given in Table~\ref{tab:spadetail}.}
\label{tab:spadeconv}
\small
\setlength{\tabcolsep}{3.5pt}\renewcommand{\arraystretch}{1.10}
\begin{tabular}{@{}crrrrrrl@{}}
\toprule
Grid & $N_{\rm pair}$ & RMSE$_{\rm DI}$ & RMSE$_{\rm SP}$ & MedAE$_{\rm DI}$ & MedAE$_{\rm SP}$ & Cat. DI/SP & $G_{\rm RMSE}$ [95\% interval] \\
\midrule
$13\times13$ & 160 & 0.3537 & 0.1810 & 0.1947 & 0.1251 & 22/3 & 1.95 [1.58--2.39] \\
$17\times17$ & 40  & 0.2906 & 0.1953 & 0.1962 & 0.1549 & 4/0  & 1.49 [1.16--1.83] \\
$17\times17$ & 80  & 0.2819 & 0.1874 & 0.1880 & 0.1304 & 8/1  & 1.50 [1.21--1.85] \\
$17\times17$ & 160 & 0.3212 & 0.1810 & 0.1786 & 0.1251 & 17/3 & \SpadeGain\ [\SpadeGainLo--\SpadeGainHi] \\
$21\times21$ & 16  & 0.3319 & 0.1951 & 0.2742 & 0.1574 & 1/0  & 1.70 [1.05--2.84] \\
$21\times21$ & 32  & 0.2946 & 0.2060 & 0.2052 & 0.1574 & 2/0  & 1.43 [1.00--2.00] \\
$21\times21$ & 48  & 0.3650 & 0.1932 & 0.2245 & 0.1498 & 7/0  & 1.89 [1.35--2.57] \\
\bottomrule
\end{tabular}
\end{table}

For the representative 17-by-17, \(\pm5\sigma_{\rm PSF}\) comparator and \(\SpadeTrials\) paired trials, direct imaging gives mass RMSE \(\SpadeDirectRMSE\%\), signed SPADE gives \(\SpadeRMSE\%\), and
\begin{equation}
G_{M,\rm RMSE}=\SpadeGain\quad[\SpadeGainLo,\SpadeGainHi],
\qquad
G_{M,\rm MedAE}=\SpadeMedAEGain\quad[\SpadeMedAEGainLo,\SpadeMedAEGainHi].
\label{eq:spadeGains}
\end{equation}
The direct/SPADE two-dimensional position RMSE changes from 0.273 to 0.169 mas and position-angle RMSE from \(11.7^\circ\) to \(7.4^\circ\). Observed catastrophic counts are \(\SpadeDirectCatCount/\SpadeTrials\) and \(\SpadeCatCount/\SpadeTrials\); every catastrophic recovery was refit from at least 12 starts and retained. The stronger qualitative result is therefore the preservation of signed sub-Rayleigh orbital information in a measurement basis for which the tested direct-image likelihood is substantially less informative.

The numerical factor remains provisional for four specific reasons. First, the strongest completed nonlinear comparator is pixel integrated; a general benchmark requires a 24-epoch event-level continuous-image likelihood evaluated on the same latent systems, priors, photon counts, and nuisance model. Second, strict tail convergence requires successive common-realization ensembles to change RMSE, MedAE, and RMSE gain by less than 5\%, and catastrophic fraction and raw 68\% coverage by less than two percentage points; none of the largest available transitions passes all five criteria. Third, the assumed broadband transfer uses internally specified crosstalk \(c=2\times10^{-4}\), residual leakage \(10^{-4}\), and modal-gain RMS \(2\times10^{-3}\); these must be replaced by an externally measured transfer matrix across the 0.9, 1.0, and \(1.1\,\mu\mathrm m\) channels. Fourth, calibration is presently represented by nuisance states and held-out perturbations; a closing experiment must allocate finite calibration photons, fit the transfer jointly with the orbit, and demonstrate held-out coverage under controlled drift. Completing those four tests would establish whether the representative gain generalizes beyond the modeled receiver.

\begin{figure}[!htbp]
\centering
\includegraphics[width=0.98\textwidth]{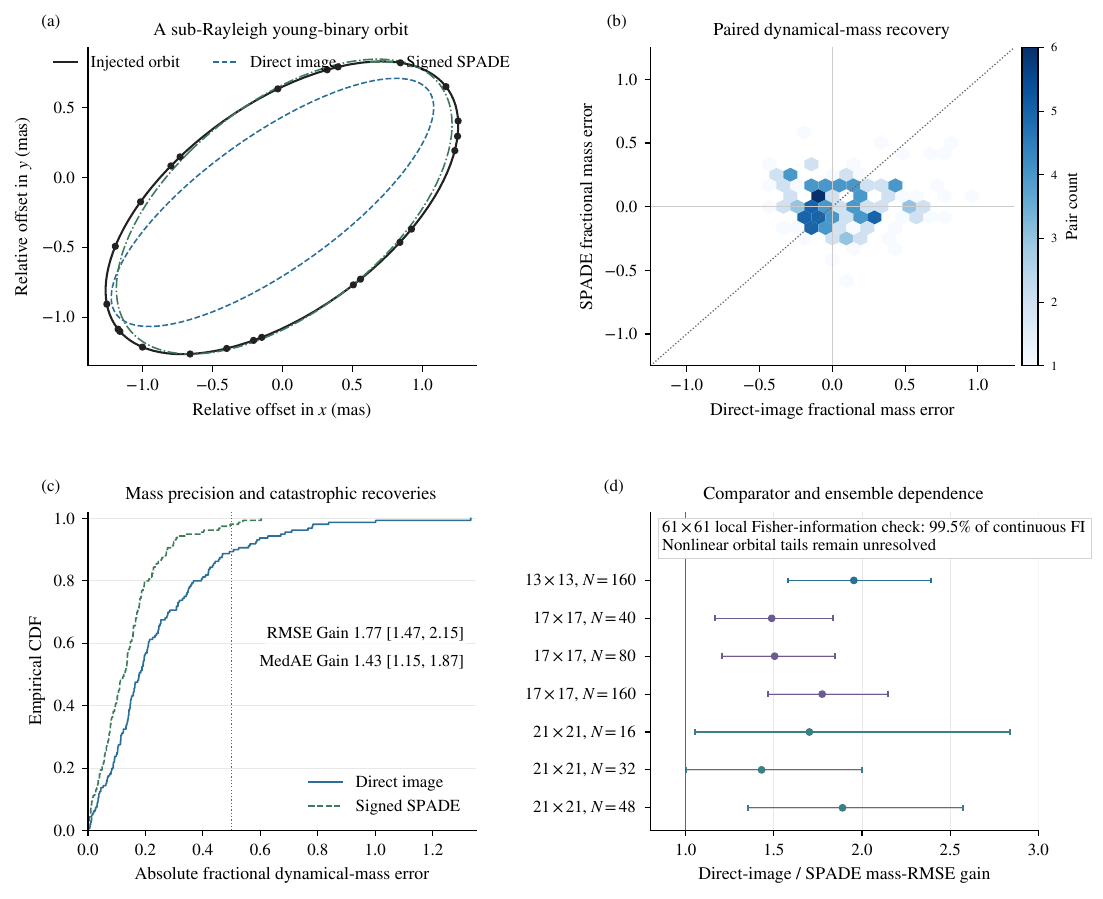}
\caption{Signed two-dimensional SPADE inference. (a) Injected orbit and one representative trial; black points mark the 24 injected epochs, and the other panels summarize the common ensemble. (b) Paired direct-image and signed-SPADE fractional dynamical-mass errors for 160 common realizations. (c) Empirical CDF of absolute mass error with the 50\% catastrophic-error criterion. (d) Dependence of mass-RMSE gain on image comparator and common-ensemble size; the \(61\times61\) annotation reports a separate local continuous-image information calculation. The 17-by-17 interval uses 10,000 paired percentile-bootstrap resamples of common trial indices with seed \SpadeBootstrapSeed; full definitions are given in Appendix~\ref{app:spade}.}
\label{fig:spadeinf}
\end{figure}

Within the matched simulator family, five-fold scaling gives nominal 68\% coverage 0.681 for both receivers and nominal 95\% coverage 0.944. These values diagnose internal consistency under the modeled transfer.

\subsection{What limits the astrophysical gain?}
Static nearest-neighbor crosstalk is tested at 0, 0.1, 0.2, 0.4, and 0.8\% in discrete 48-pair experiments.  At 0.4\%, the median mass-RMSE gain is 1.36 with interval 0.95--1.88; a gain above unity is not established at 95\% confidence.  Separate held-out mechanisms vary origin error in \(\sigma_{\rm PSF}\), sorter angle in milliradians, modal-gain RMS, and crosstalk drift.  The combined stability case uses \(0.0025\sigma_{\rm PSF}\) origin RMS, 0.5 mrad angle RMS, \(10^{-3}\) modal-gain RMS, and \(2\times10^{-4}\) crosstalk-drift RMS.  It yields median gain \SpadeDriftGain\ (\SpadeDriftGainLo--\SpadeDriftGainHi) and empirical 68\% coverage \SpadeDriftCoverage\% (\SpadeDriftCoveredCount/48; Wilson 95\% interval \SpadeDriftCoverageLo--\SpadeDriftCoverageHi\%). Under this combined-drift stress case, a SPADE advantage is not established and the nominal intervals under-cover severely. The experiment does not establish that the point-estimate RMSE benefit vanishes.

For modal gain or crosstalk state \(x_{e,k}\), the stress model is
\begin{equation}
\operatorname{Cov}[\delta x_{e,k},\delta x_{e',k'}]
=\sigma_x^2\rho_t^{|e-e'|}
[\rho_\lambda+(1-\rho_\lambda)\delta_{kk'}],
\label{eq:spadedrift}
\end{equation}
where \(\rho_t\) corresponds to a 45-d correlation time and \(\rho_\lambda=0.8\).  The tested points are finite ensembles; nonmonotonic point estimates with broad intervals are not interpolated as a smooth tolerance curve.

\begin{figure}[!htbp]
\centering
\includegraphics[width=0.98\textwidth]{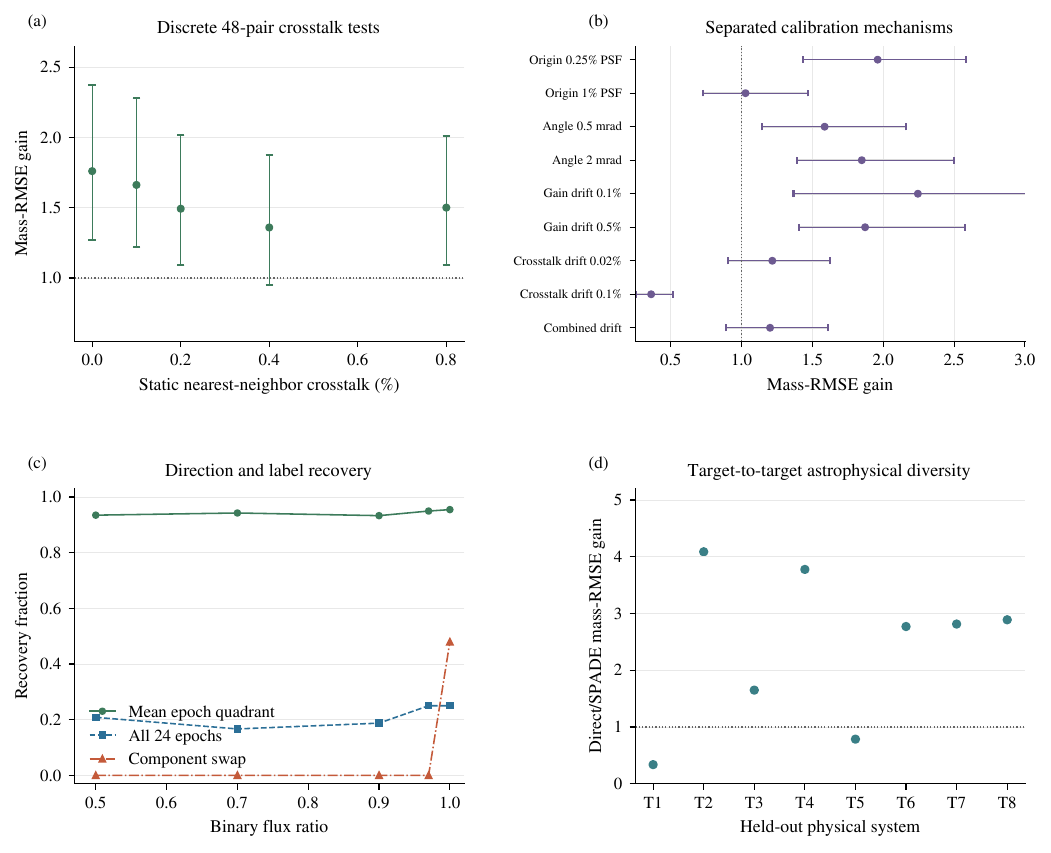}
\caption{SPADE calibration and target-diversity stress tests.  (a) Discrete 48-pair static nearest-neighbor crosstalk experiments; points are median RMSE gains and bars are bootstrap 95\% intervals, with unity shown lightly.  (b) Separately varied calibration mechanisms in physical units and one combined case.  (c) Signed-direction and label recovery versus binary flux ratio; equal-flux systems are physically ambiguous under component exchange.  (d) Direct/SPADE mass-RMSE gain for eight held-out Taurus-like physical systems.  The cluster-bootstrap population interval resamples systems, not repeated photon realizations.}
\label{fig:spadecal}
\end{figure}

At equal flux, 23 of 48 trials select the exchanged component label, exposing the intrinsic symmetry rather than concealing it with a prior.  The out-of-family population spans eight independent Taurus-like systems with masses 0.06--\(0.30M_\odot\), distances 120--170 pc, flux ratios 0.30--0.90, eccentricities 0.05--0.60, and inclinations \(25^\circ\)--\(80^\circ\).  Four photon realizations are retained per system.  Across the eight held-out physical systems, the population mass-RMSE gain is 1.84 with interval 1.14--3.00; the large target-to-target spread shows that orbit geometry and source properties remain astrophysically important.

To translate mass precision into an illustrative astrophysical consequence, consider equal-prior Gaussian predictions \(M_1=0.10M_\odot\) and \(M_2=0.12M_\odot\), a common fractional one-sigma precision \(p\) evaluated at \(\bar M=0.11M_\odot\), and the midpoint decision boundary. The correct-selection probability is
\begin{equation}
P_{\rm correct}=\Phi\!\left(\frac{M_2-M_1}{2p\bar M}\right),
\label{eq:pmsselect}
\end{equation}
which gives 72.8\%, 81.8\%, and 90.3\% for \(p=15\%\), 10\%, and 7\%, respectively. Rounded to whole percentages, the science-facing sequence is 73\%, 82\%, and 90\%. This calculation isolates the leverage of dynamical-mass precision; a named-system analysis would additionally infer age, magnetic inflation, accretion history, radii, and atmosphere-model uncertainty~\cite{Rizzuto2020YoungBinaries,Baraffe2015,Feiden2016}.

\subsection{Interpretation}
The physical result is a change in the parameter-bearing measurement basis: first-order modal counts retain signed separation information that the image-plane likelihood loses near the Rayleigh limit. Carrying that information through a Keplerian orbit makes close young binaries the strongest positive opportunity in this comparison, because improved mass precision can change the discrimination among pre-main-sequence predictions. Calibration drift, component labeling, crosstalk, and target-specific orbit geometry then become the practical bottlenecks, as quantified in Fig.~\ref{fig:spadecal}. The focused closure program in the preceding subsection defines the experiment and analysis needed to convert that qualitative information advantage into a more general numerical gain.

\section{Receiver-level opportunities for exoplanet imaging and coherent astronomy}
\label{sec:maps}
The next comparisons stop at receiver level.  They answer whether a deeper instrument and astrophysical analysis is warranted; they do not assign a target-level mission gain.

\subsection{Modal coronagraphy}
\begin{measurementstatement}
\msitem{Astrophysical observable}{Planet and stellar photon rates after high-contrast spatial filtering.}
\msitem{Limiting classical effect}{The throughput--leakage balance near the inner working angle: suppressing the star while retaining enough planet photons.}
\msitem{Quantum intervention}{A quantum-estimation-informed projection that rejects the dominant on-axis stellar mode while retaining part of an off-axis planet field.}
\msitem{Science-facing metric}{Poisson integration time or planet-amplitude standard deviation.}
\msitem{Required classical comparator}{A conventional coronagraph with the same pupil, separation, source contrast, background, and declared throughput/leakage frontier.}
\msitem{Conditions for astrophysical gain}{Finite-star leakage, aberration, chromaticity, mode mismatch, crosstalk, and planet throughput must remain below the conventional frontier.}
\msitem{Present scientific status}{Analytical receiver-level requirement map with supporting laboratory literature.}
\end{measurementstatement}

An on-axis unresolved star is concentrated in a dominant spatial mode.  A modal coronagraph projects that mode away; an off-axis planet is not perfectly aligned with the rejected mode and retains finite throughput.  Finite stellar diameter populates orthogonal modes, while wavefront errors, pointing, chromaticity, and transfer-matrix crosstalk reintroduce stellar photons.  The receiver therefore trades planet throughput against stellar rejection rather than eliminating leakage absolutely.

For contrast \(C\), planet throughput \(T_p\), residual stellar/instrumental leakage \(L\), and background \(B\), the time for fixed planet-amplitude signal-to-noise scales as
\begin{equation}
t\propto\frac{CT_p+L+B}{(CT_p)^2}.
\label{eq:coro}
\end{equation}
The reference accounting uses \(C=10^{-5}\), separation \(0.7\lambda/D\), stellar diameter \(0.014\lambda/D\), modal planet throughput 0.501, finite-star leakage \(3.47\times10^{-5}\), crosstalk \(10^{-4}\), and background \(10^{-6}\) of the stellar rate.  The conventional reference uses throughput 0.30 and calibrated leakage \(10^{-4}\).  Table~\ref{tab:coronaaccounting} exposes every term entering the nominal ratio.  For Poisson amplitude estimation, \(t\propto n/p^2\), where \(p\) is the accepted planet rate and \(n\) the total accepted rate; the modal-to-conventional integration-time ratio is 0.486, corresponding to speed ratio \(\CoronagraphSpeed\) and standard-deviation gain \(\CoronagraphPrecision\).  At this nominal point the modal receiver accepts more total nonplanet counts than the conventional reference, \(1.36\times10^{-4}\) versus \(1.01\times10^{-4}\) of the unattenuated stellar rate, but it also accepts more planet photons.  The favorable result is therefore a combined Poisson throughput--leakage trade, not an unconditional reduction of total leakage.

\begin{table}[tb]
\caption{Normalized modal-coronagraph throughput--leakage accounting.  Rates are fractions of the unattenuated stellar rate.  ``Nonplanet'' is finite-star plus instrumental/crosstalk leakage plus background.  The nominal speed ratio is conditional on both rows and follows from the complete Poisson balance, not from lower total leakage alone.}
\label{tab:coronaaccounting}
\small
\setlength{\tabcolsep}{3.2pt}\renewcommand{\arraystretch}{1.12}
\begin{tabular}{@{}lrrrrrrrr@{}}
\toprule
Receiver & \(T_p\) & Planet & Finite star & Inst./cross & Background & Nonplanet & Total & \(t/t_{\rm conv}\) \\
\midrule
Modal & 0.501 & \(5.01\times10^{-6}\) & \(3.47\times10^{-5}\) & \(1.00\times10^{-4}\) & \(1.00\times10^{-6}\) & \(1.36\times10^{-4}\) & \(1.41\times10^{-4}\) & 0.486 \\
Conventional & 0.300 & \(3.00\times10^{-6}\) & 0 & \(1.00\times10^{-4}\) & \(1.00\times10^{-6}\) & \(1.01\times10^{-4}\) & \(1.04\times10^{-4}\) & 1.000 \\
\bottomrule
\end{tabular}
\end{table}

A conditional classical frontier is also evaluated by varying conventional planet throughput from 0.20 to 0.60 and residual stellar leakage from \(10^{-6}\) to \(10^{-3}\), with the planet contrast and background held fixed.  The modal receiver is favorable only on the portion of that throughput--leakage surface where \(n_{\rm conv}/p_{\rm conv}^2>n_{\rm modal}/p_{\rm modal}^2\).  Consequently the nominal \(2.06\) speed ratio is an illustrative receiver comparison, while the frontier in Fig.~\ref{fig:maps} is the more transferable requirement.  This result is compatible with modal-coronagraph laboratory demonstrations~\cite{Deshler2025Experiment,Deshler2026Coronagraph}; a target-level inference claim requires an instrument-specific, time-dependent wavefront and exozodiacal model.

\subsection{Quantum-enhanced coherent reception}
\begin{measurementstatement}
\msitem{Astrophysical observable}{Field quadratures or complex amplitude in a coherent spectro-interferometric receiver.}
\msitem{Limiting classical effect}{The image-band vacuum contribution of heterodyne detection.}
\msitem{Quantum intervention}{Two-mode squeezing reduces the image-band variance entering the coherent estimator.}
\msitem{Science-facing metric}{Fisher-information and standard-deviation gain over classical heterodyne; separately, information relative to direct detection.}
\msitem{Required classical comparator}{Optimized classical heterodyne for the first comparison and ideal direct detection for the second, with each denominator stated separately.}
\msitem{Conditions for astrophysical gain}{Optical loss, local-oscillator/reference distribution, weak-mode occupation, background, bandwidth, and array-wide phase stability.}
\msitem{Present scientific status}{Analytical receiver-level map with laboratory demonstrations of the receiver mechanism.}
\end{measurementstatement}

Classical heterodyne reception mixes the astronomical field with a local oscillator and estimates both field quadratures.  The unused image band contributes vacuum fluctuations, producing a noise penalty independent of source brightness.  Two-mode squeezed reception can correlate or suppress that contribution, reducing the variance of the complex-amplitude estimator.  It does not increase the astronomical occupation number and therefore need not outperform direct photon counting in the weak-source regime.

With background occupation \(n_b\) and image-band variance ratio \(V_{\rm img}=10^{-3.5/10}=0.447\),
\begin{equation}
G_{\rm FI}^{q/{\rm het}}=\frac{1+n_b}{V_{\rm img}+n_b},
\qquad
G_{\sigma}^{q/{\rm het}}=\sqrt{G_{\rm FI}^{q/{\rm het}}}.
\label{eq:qhet}
\end{equation}
At \(n_b=10^{-3}\), \(G_{\rm FI}=\QHetFIGain\) and \(G_\sigma=\QHetSdGain\), consistent with demonstrated receiver-level noise suppression~\cite{Gould2024Heterodyne,Anai2024Heterodyne}.  In the weak-occupation domain \(\epsilon,n_b\le10^{-2}\), the reference comparison to ideal direct detection is
\begin{equation}
\frac{F_{\rm qhet}}{F_{\rm dir}}\simeq
\epsilon\frac{1+n_b}{V_{\rm img}+n_b}.
\label{eq:qhetdirect}
\end{equation}
For a 20-Jy source at 10 \(\mu\)m, a 2-m collector, and 20\% throughput, \(\epsilon\simeq6.3\times10^{-6}\) and \(F_{\rm qhet}/F_{\rm dir}=\QHetDirectFI\).  Quantum enhancement improves heterodyne but does not remove the weak-occupation disadvantage relative to direct detection.  Coherent reception can still be selected for spectral resolution, frequency stability, or field-transport architecture.  Quantum-correlated dual-comb spectroscopy is a separate demonstrated positive for active spectroscopy~\cite{Herman2025,Wan2025,Hariri2025}.

\subsection{Astrophysical niches and break-even conditions}
\begin{figure}[!htbp]
\centering
\includegraphics[width=0.98\textwidth]{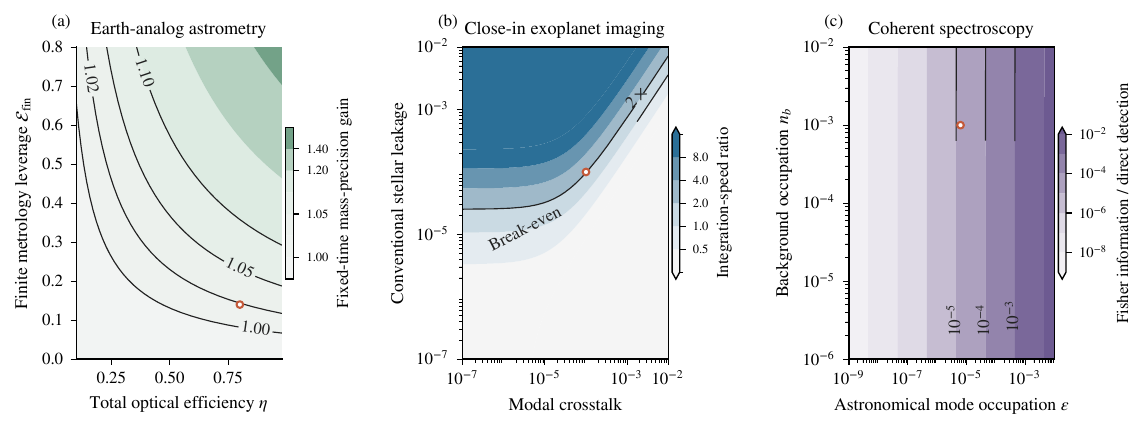}
\caption{Analytical receiver-level requirement maps.  (a) Fixed-time squeezed-metrology precision gain versus total optical efficiency and effective finite-intervention sensitivity, with contours at unity and selected gains.  (b) Modal-coronagraph precision gain versus modal crosstalk and conventional leakage for the photon-rate accounting in Eq.~\eqref{eq:coro}; the dark contour is break-even and the marker is the nominal comparison.  (c) Quantum-heterodyne Fisher information relative to direct detection in the weak-occupation domain of Eq.~\eqref{eq:qhetdirect}; muted vertical bands indicate the classical-heterodyne and quantum-enhanced-heterodyne regions of the reference occupation slice, and the marker is the 20-Jy mid-infrared example.  These panels establish receiver-level break-even requirements rather than astrophysical inference.}
\label{fig:maps}
\end{figure}

Figure~\ref{fig:maps} generalizes the two receiver opportunities and the squeezed-metrology boundary.  The advantage survives where the quantum-modified term remains a substantial part of the receiver information and disappears when throughput, leakage, weak occupation, or an uncoupled covariance term dominates.  The next evidence level is an instrument-specific likelihood including wavefront control for the coronagraph and local-oscillator/reference distribution for coherent reception.  Other mechanisms occupy distinct niches.  Two-photon optical-coherence tomography and dispersion-canceling interference improve active delay and dispersion measurements rather than passive stellar astrometry~\cite{Abouraddy2002,Okano2013}.  Induced-coherence imaging provides wavelength access, while classical phase-sensitive analogues require a matched comparator~\cite{Lemos2014,Shapiro2015Undetected}.  Cross-comb approaches address active spectroscopy and frequency conversion~\cite{Liu2023CrossComb}.  Two-photon amplitude astrometry and intensity interferometry avoid ordinary coherent field transport but trade against accepted-pair or coincidence rate~\cite{Stankus2022TwoPhoton,Chen2023TwoPhoton,Crawford2023TwoPhoton,Bojer2022}.

\section{Entanglement-assisted long-baseline interferometry: opportunity and present limits}
\label{sec:network}

\begin{measurementstatement}
\msitem{Astrophysical observable}{First-order mutual coherence and source parameters encoded in the complex visibility.}
\msitem{Limiting classical effect}{Coherent transport of the astronomical field over a long physical baseline.}
\msitem{Quantum intervention}{Ancillary entanglement and memories mediate a nonlocal two-output measurement.}
\msitem{Science-facing metric}{Phase- or visibility-amplitude Fisher information per completed information-bearing operation.}
\msitem{Required classical comparator}{The finite-aperture Fresnel coherent relay under the same source, wavelength, bandwidth, and observing time.}
\msitem{Conditions for astrophysical gain}{A declared phase schedule, nonzero source visibility and derivative, completed-operation yield, false-event control, pair delivery, memory, and phase-reference closure.}
\msitem{Present scientific status}{Receiver/protocol-level negative control. The declared model supports the event likelihood and unencoded operation model, but does not define a diameter-crossover requirement.}
\end{measurementstatement}

\subsection{Physical mechanism and phase-dependent information}
A nonlocal receiver attempts to measure the single-photon mutual coherence without transporting the astronomical photon to a common combiner. Let the true accepted-event rate be $s$, false accepted-event rate $f$, source visibility $V_{\rm src}(\boldsymbol\theta)$, receiver contrast $v_q$, and controlled phase $\phi$. The output rates are
\begin{equation}
r_\pm=\frac{s}{2}\left[1\pm v_qV_{\rm src}(\boldsymbol\theta)\cos\phi\right]+\frac{f}{2}.
\label{eq:networkrates}
\end{equation}
For a parameter that enters through interferometric phase, the Poisson Fisher rate is
\begin{equation}
\dot F_{\phi\phi}=\frac{s^2(v_qV_{\rm src})^2(s+f)\sin^2\phi}
{(s+f)^2-s^2(v_qV_{\rm src})^2\cos^2\phi}.
\label{eq:networkphasefi}
\end{equation}
For a parameter $\theta$ that changes only visibility amplitude,
\begin{equation}
\dot F_{\theta\theta}=\frac{s^2v_q^2(\partial_\theta V_{\rm src})^2(s+f)\cos^2\phi}
{(s+f)^2-s^2v_q^2V_{\rm src}^2\cos^2\phi}.
\label{eq:networkampfi}
\end{equation}
Thus quadrature, $\phi=\pi/2$, maximizes phase information but gives exactly zero amplitude information. Diameter, limb darkening, wavelength scale, and instrumental contrast therefore require amplitude-sensitive phase settings and, when phase is also unknown, a predeclared phase-diverse schedule. A finite $f/s=O(1)$ imposes a finite information penalty; if $f$ remains finite as $s\rightarrow0$, the information returns to quadratic weak-signal scaling.

\subsection{Which source parameters are actually measurable?}
The declared configuration contains no phase-step schedule, and the available implementation evaluates phase information at quadrature. Because the diameter, limb-darkening, wavelength-scale, and contrast parameters enter the visibility amplitude, the model cannot generate a valid diameter-information boundary. Accordingly this paper reports no network/classical diameter crossover, logical-yield requirement, or pair and memory requirement derived from such a crossover. Table~\ref{tab:networkcheck} records the equation-to-implementation closure test.

% Inlined table definition.
\small
\setlength{\tabcolsep}{4pt}\renewcommand{\arraystretch}{1.10}
\begin{longtblr}[caption={Nonlocal-likelihood consistency calculation. The declared physical model contains no phase-stepping schedule for visibility-amplitude inference.},label={tab:networkcheck}]{width=\textwidth,colspec={X[24,l] X[14,c] X[54,l]},rowhead=1}
\toprule
Quantity & Status & Result\\\midrule
Two-output Poisson rates & established & The rate model is mathematically well defined.\\
Phase Fisher information & established & The analytic expression agrees with finite differences and is maximal at $\phi=\pi/2$.\\
Visibility-amplitude information at quadrature & zero & It is proportional to $\cos^2\phi$ and vanishes at $\phi=\pi/2$.\\
Declared phase schedule & unresolved & No phase-step array is present in the declared physical model.\\
Diameter-information boundary & unsupported & A phase-diverse likelihood is required before a source-diameter crossover can be calculated.\\
Sirius-B coherence at 1355 km & negative control & $\langle|V|\rangle=6.17\times10^{-5}$ and $V_{\rm RMS}=6.88\times10^{-5}$; the uniform-disk-equivalent first null is 8.81 km.\\
\bottomrule
\end{longtblr}
\normalsize

The conservative scientific interpretation is a phase-diverse receiver likelihood evaluated with the same source parameters, nuisance priors, baselines, and channels as the coherent comparator. A suitable fixed design would include amplitude-sensitive settings near $0$ and $\pi$ and quadrature settings near $\pi/2$ and $3\pi/2$, selected before examining the architecture comparison. That calculation remains future work; no replacement crossover is inferred here.

\subsection{Source coherence and receiver completion}
\begin{figure}[!htbp]
\centering
\includegraphics[width=0.98\textwidth]{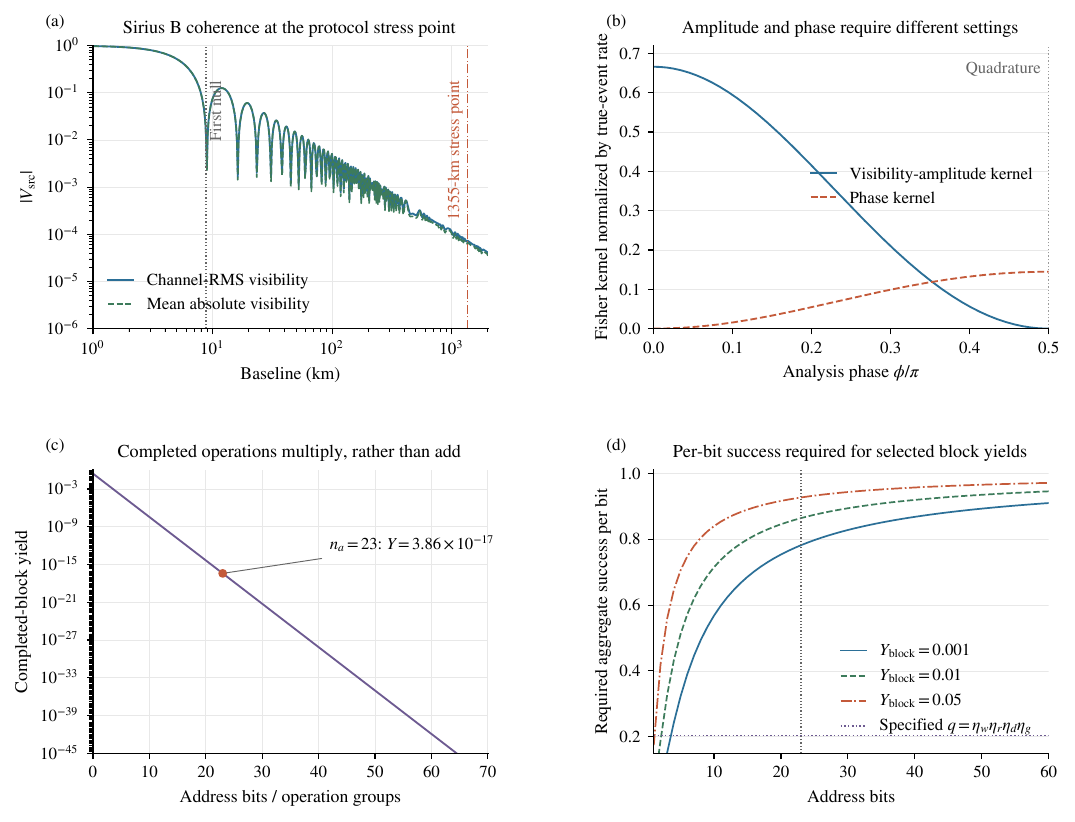}
\caption{Nonlocal-reception likelihood consistency and protocol boundary. (a) Weighted mean absolute visibility and RMS visibility of the Sirius-B channel model versus baseline at $1\,\mu$m; the dotted line is the uniform-disk-equivalent first null and the dashed line is the 1355-km protocol stress point. (b) Normalized kernels $K_\phi=\dot F_{\phi\phi}/s$ and $K_{\rm amp}=\dot F_{\theta\theta}/[s(\partial_\theta V_{\rm src})^2]$ for $f/s=0.1055$, $v_q=0.8$, and representative $V=0.5$; quadrature maximizes phase information and gives zero amplitude information. (c) Operation-counted unencoded block yield versus address bits, with the declared 23-bit point. (d) Aggregate per-bit success $q=(Y_{\rm block}/0.286)^{1/n}$ required for selected hypothetical block yields; these are protocol requirements, not astrophysical crossover predictions.}
\label{fig:network}
\end{figure}

The protocol stress test retains the Sirius-B source model because it exposes two independent limitations.  The source is a 25,000-K, $0.0081R_\odot$ disk at 2.637 pc, corresponding to an angular diameter of $28.57\,\mu$as, in a 10-nm band centered at $1\,\mu$m.  The adopted linear limb-darkening coefficient is $u=0.15$, with intensity law
\[
I(\mu)=I(1)\,[1-u(1-\mu)] .
\]
For $x=\pi B\theta/\lambda$, the normalized monochromatic visibility is
\[
V(x)=\frac{(1-u)J_1(x)/x+u(\sin x-x\cos x)/x^3}{(1-u)/2+u/3},
\qquad V(0)=1.
\]
The spectral grid contains 300 equal-frequency channels spanning $0.995$--$1.005\,\mu$m.  Each channel has a uniform top-hat response, and the stored band summaries use uniform channel weights $w_i=1$, so the normalized weights are $w_i/\sum_j w_j=1/300$.  The total incident source rate is $2.28\times10^4\,\mathrm{s^{-1}}$ per 0.5-m telescope; the separate radiometric background allocation is 5\% of that rate.  Figure~\ref{fig:network}(b) uses the independently declared accepted false-to-true event ratio $f/s=0.1055$ from the network-likelihood configuration; the available records do not derive this ratio from the 5\% radiometric background alone.  For channel weights $w_i$, we distinguish
\begin{equation}
\langle|V|\rangle=\frac{\sum_iw_i|V_i|}{\sum_iw_i},\qquad
V_{\rm RMS}=\left(\frac{\sum_iw_i|V_i|^2}{\sum_iw_i}\right)^{1/2}.
\label{eq:networkvismoments}
\end{equation}
The uniform-disk-equivalent first null is $B_1\simeq1.22\lambda/\theta=\NetworkFirstNull$ km.  At the 1355-km protocol stress point the channel summaries are $\langle|V|\rangle=\NetworkMeanAbsV$ and $V_{\rm RMS}=\NetworkRmsV$.  These small correlated-flux amplitudes identify an over-resolved source, but small $|V|$ alone does not prove negligible diameter information: $\partial V/\partial\theta$, the phase schedule, spectral response, calibration, and parameter degeneracies also enter the likelihood.  Because the declared model lacks the required phase-diverse schedule, Sirius B is used here only as a source-coherence and protocol negative control.

The source-plus-background occupation in channel $i$ defines the block size
\begin{equation}
\epsilon_i=\frac{2(r_{s,i}+r_{b,i})}{\Delta\nu_i},\qquad
M_i=\frac{\mu}{\epsilon_i},\qquad n_{a,i}=\lceil\log_2M_i\rceil,
\label{eq:address}
\end{equation}
with $\mu=0.1$ and $n_{a,i}=23$ for all channels. Multiphoton and background-only accepted blocks contribute to $f$ unless a physical number-discrimination and rejection module is supplied. With one aggregated write, read, detection, and gate operation per address bit, the specified unencoded yield is
\begin{equation}
Y_{\rm unc}=\eta_{\rm cap}\eta_{\rm dec}d(\eta_w\eta_r\eta_d\eta_g)^{23}=\NetworkUncYield,
\label{eq:yunc}
\end{equation}
using $(\eta_{\rm cap},\eta_{\rm dec},d)=(0.65,0.80,0.55)$ and $(\eta_w,\eta_r,\eta_d,\eta_g)=(0.55,0.55,0.90,0.75)$. A station-by-station implementation would compound more operations and reduce the yield further. This reproducible result rules out only the specified unencoded operation model.

\subsubsection{Interpretation}
The physical nonlocal measurement principle remains well motivated, but the benchmark reaches only a protocol/source negative control. At this stress baseline Sirius B is highly over-resolved, with very small band-averaged correlated-flux amplitude. The next evidence level requires a predeclared phase-diverse likelihood, a compact source with measurable correlated flux, and a complete encoded receiver that closes pair, memory, phase-reference, false-event, and retry resources against the optimized classical link.

\section{When a quantum gain does not improve the astrophysics}
\label{sec:controls}

\subsection{Squeezed nuller sensing}
\begin{measurementstatement}
\msitem{Astrophysical observable}{Planet continuum amplitude in a four-collector mid-infrared null output.}
\msitem{Limiting classical effect}{Closed-loop piston disturbance, actuator residual, geometric/instrumental leakage, and diffuse/thermal backgrounds.}
\msitem{Quantum intervention}{Six-decibel squeezing reduces the phase-sensor noise power spectral density in an independently optimized controller.}
\msitem{Science-facing metric}{Closed-loop residual OPD spectrum and marginalized planet-amplitude precision.}
\msitem{Required classical comparator}{An independently optimized unsqueezed controller under the same plant, stability margins, actuator constraints, scene, exposure, and nuisance priors.}
\msitem{Conditions for astrophysical gain}{Sensor noise must contribute materially to the closed-loop residual and then to the planet likelihood.}
\msitem{Present scientific status}{Instrument-level finite-ensemble negative control using a marginalized planet-amplitude precision proxy under a reduced-order model.}
\end{measurementstatement}

A nuller uses destructive interference to reject the on-axis star.  Improving the phase sensor can only help if sensor noise propagates through the feedback loop into a residual OPD term that is important relative to disturbance, actuator noise, stellar leakage, and astrophysical backgrounds.  Squeezing the control sensor does not directly suppress geometric leakage, amplitude mismatch, exozodiacal emission, or telescope thermal photons.

The four collectors are at \((-3/2,-1/2,1/2,3/2)B\) with weights \((1,-1,-1,1)/2\), giving fourth-order near-axis intensity transmission.  The reference benchmark has four 2-m collectors, \(B=20\) m, 8--13 \(\mu\)m, a Sun--Earth analog at 10 pc, 10-d integration, local and three-zodi exozodiacal backgrounds, a 40-K telescope with 5\% emissivity, detector background, geometric stellar leakage, and instrument leakage from piston, 0.1\% amplitude mismatch, 1-mas tip/tilt, chromatic phase, polarization mismatch, and correlated piston modes.

The plant and controller are
\begin{equation}
P(s)=\frac{e^{-s\tau_d}}{1+s/(2\pi f_p)},
\qquad
C(s)=K\frac{1+s/(2\pi f_z)}{s[1+s/(2\pi f_h)]},
\label{eq:nullcontrol}
\end{equation}
with \(\tau_d=5\) ms and \(f_p=80\) Hz.  For disturbance, sensor, and actuator one-sided power spectral densities \(S_d,S_n,S_a\),
\begin{equation}
S_x^{\rm res}=|S|^2S_d+|T|^2S_n+|PS|^2S_a,
\quad S=(1+PC)^{-1},\quad T=PC(1+PC)^{-1}.
\label{eq:nullpsd}
\end{equation}
Classical and squeezed controllers are optimized independently with phase margin at least \(35^\circ\), gain margin at least 2, and command RMS below 80 nm.  Planet amplitude, local-zodiacal, exozodiacal, thermal, and stellar-leak calibration factors are fit jointly.

The plotted quantum-induced precision change is
\begin{equation}
\Delta_q=10^6\frac{\sigma_{\rm cl}-\sigma_q}{\sigma_{\rm cl}}
\quad \text{ppm},
\label{eq:nullmetric}
\end{equation}
so positive values mean smaller uncertainty with the squeezed sensor.  Across 180 held-out plant/background realizations, the sampled 95th percentile is \(\NullerPppm\) ppm with bootstrap interval 9.6--11.6 ppm.  It is an empirical percentile, not a mathematical bound.

\begin{figure}[!htbp]
\centering
\includegraphics[width=0.98\textwidth]{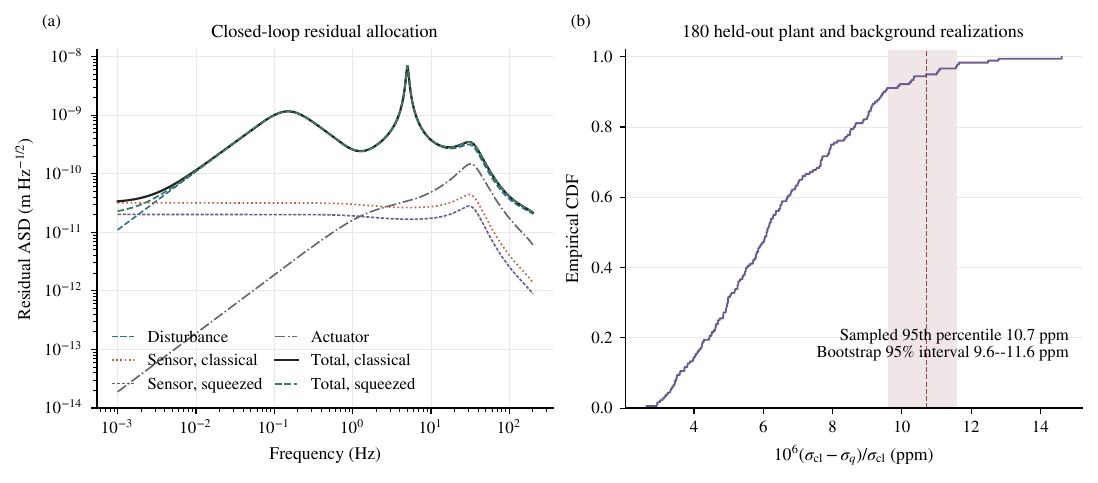}
\caption{Finite-ensemble nuller negative control.  (a) Closed-loop amplitude spectral densities (ASDs): disturbance, classical and squeezed sensor contributions, actuator contribution, and total residuals.  The total is emphasized with a neutral curve; the sensor terms remain subdominant over the information-bearing band.  (b) Empirical CDF of \(\Delta_q=10^6(\sigma_{\rm cl}-\sigma_q)/\sigma_{\rm cl}\) for 180 held-out plant/background realizations.  The dashed line is the sampled 95th percentile and the band its bootstrap 95\% interval.}
\label{fig:nuller}
\end{figure}

\subsubsection{Interpretation}
The squeezed receiver lowers the sensor term, but disturbance and actuator residuals dominate the loop and diffuse backgrounds dominate the science likelihood.  The new bottleneck is therefore unchanged by the quantum intervention.  The present evidence establishes a finite-ensemble negative control for this plant and background allocation; a next experiment would need a configuration in which sensor noise is intentionally raised into the final likelihood budget or a hardware-in-the-loop null spectrum with independently measured residual terms.

\subsection{Canonical NOON prepared probe}
\begin{measurementstatement}
\msitem{Astrophysical observable}{None directly; this is a local active phase-estimation control.}
\msitem{Limiting classical effect}{Shot noise of a coherent phase probe.}
\msitem{Quantum intervention}{An \(N\)-photon path-entangled NOON state accumulates \(N\phi\).}
\msitem{Science-facing metric}{Wall-clock standard-deviation gain at fixed attempted input photons and declared ancillary overhead.}
\msitem{Required classical comparator}{A coherent probe with the same attempted photons, detector performance, path, and elapsed time.}
\msitem{Conditions for astrophysical gain}{Preparation and heralding, all-photon survival, intrinsic visibility, phase diffusion, duty, detector efficiency, and retry/discard costs.}
\msitem{Present scientific status}{Analytical receiver-level comparison of one canonical implementation.}
\end{measurementstatement}

The ideal state \((|N,0\rangle+|0,N\rangle)/\sqrt2\) has an \(N\)-fold phase response and lossless \(N^2\) Fisher information per successful state.  Accepted-event normalization can conceal the dominant architecture costs: a failed preparation, lost photon, rejected herald, or discarded trial consumes time and attempted photons even though it contributes no accepted phase sample.  At fixed attempted photons and wall-clock time, the modeled gain is
\begin{equation}
G_N=V_N e^{-N^2\sigma_\phi^2/2}
\left[
\frac{P_{\rm prep}P_{\rm herald}d}{\kappa}
N\eta^{N-1}
\right]^{1/2},
\label{eq:noon}
\end{equation}
where \(\kappa\) is coherent-equivalent ancillary and attempt overhead.  The declared accounting uses \(N=1,\ldots,20\), \(P_{\rm prep}=0.05\), \(P_{\rm herald}=1\) already included in completed preparation, \(d=0.70\), \(V_N=0.95\), \(\sigma_\phi=0.03\) rad, \(\eta=0.80\), and \(\kappa=1\).  The maximum is \(G_N=\NoonMaxGain\) at \(N=4\).  The modeled canonical implementation is therefore inferior to the coherent probe under this attempted-resource accounting.  Loss-optimized, adaptive, or encoded non-Gaussian probes require separate architectures and are not classified by Eq.~\eqref{eq:noon}.

\subsubsection{Interpretation}
The ideal phase response is physically improved, but preparation probability, all-photon survival, phase diffusion, and attempt overhead reduce the wall-clock information rate below the coherent reference.  The bottleneck therefore migrates from coherent-probe shot noise to source and survival throughput.  The justified claim is limited to the declared canonical implementation; the next evidence level requires a specified loss-optimized or encoded probe with measured preparation and accepted-information rates under the same attempted-photon and elapsed-time denominator.

\section{Astrophysical opportunity map and decisive demonstrations}
\label{sec:synthesis}
The calculations point to three different scientific dispositions rather than a single ranking of quantum technologies. Signed spatial-mode detection offers the clearest route to a qualitatively different astronomical measurement because it preserves sub-Rayleigh separation information that is otherwise lost. Squeezed internal metrology improves a relevant observable, but in the Earth-analog benchmark the final mass precision is controlled mainly by nonquantum covariance and observing geometry. Modal coronagraphy remains a conditional opportunity whose value is set by the planet-throughput and stellar-leakage frontier. The remaining architectures are valuable chiefly because they identify the conditions that must change before new astrophysics becomes possible. Table~\ref{tab:astro_summary} condenses the science conclusion and decisive closure test for each architecture.

\small
\setlength{\tabcolsep}{3.5pt}\renewcommand{\arraystretch}{1.12}
\begin{longtblr}[caption={Astrophysical interpretation of the comparative benchmarks. The final column identifies the experiment or calculation needed to convert the present result into a stronger astronomy claim.},label={tab:astro_summary}]{width=\textwidth,colspec={X[14,l] X[19,l] X[32,l] X[30,l]},rowhead=1}
\toprule
Technique & Science observable & Present astrophysical conclusion & Decisive next demonstration \\\midrule
Squeezed internal metrology & Earth-analog astrometric mass & The receiver gain survives to the posterior, but only as a \EarthFixedGain\ fixed-time mass-precision factor; larger leverage comes from mission duration, baseline, cadence, zero point, and stellar-photocenter control & Hardware-in-the-loop astrometric calibration with measured coupling, duty, backscatter, zero-point sharing, and recovered angular signal \\
Signed SPADE & Close-binary orbit and dynamical mass & Signed SPADE preserves sub-Rayleigh orbital information in the tested comparator. The representative mass-RMSE gain is \SpadeGain; illustrative 0.10 versus \(0.12M_\odot\) model selection is 73\%, 82\%, and 90\% at 15\%, 10\%, and 7\% mass precision & Execute the focused comparator-and-calibration closure program defined in Sec.~\ref{sec:spade}, followed by controlled orbit recovery \\
Modal coronagraphy & Planet photon rate near the inner working angle & The nominal receiver can reduce exposure time only on a specified throughput--leakage surface; no target-level spectrum or detection yield is yet established & Simultaneous throughput, finite-star leakage, crosstalk, wavefront, and exozodiacal-background test propagated to a planet spectrum or localization posterior \\
Quantum heterodyne & Complex amplitude and high-resolution spectrum & It improves classical heterodyne, but direct detection remains more informative for the selected weak optical mode; its astrophysical niche is therefore architectural or spectroscopic rather than raw photon sensitivity & Common-source comparison of quantum heterodyne, classical heterodyne, and direct detection with the same bandwidth, detector, local oscillator, phase reference, and elapsed time \\
Entanglement-assisted reception & Long-baseline complex visibility & The present stress case cannot measure visibility amplitude at quadrature, Sirius B is highly over-resolved, and the unencoded completion yield is negligible; no diameter science follows from this configuration & Predeclared phase-diverse likelihood on a source with useful correlated flux, plus complete encoded pair, memory, false-event, retry, and phase-reference accounting \\
Squeezed null sensing & Mid-infrared null depth and planet amplitude & Lower sensor noise does not change the final science precision when disturbance, actuator residual, stellar leakage, and diffuse backgrounds dominate & Nuller configuration in which sensor noise is a measured material part of the closed-loop residual and planet likelihood \\
Prepared NOON probe & Active phase reference & The canonical attempted-photon implementation is inferior to a coherent probe and is not propagated to a passive-astronomy observable & Loss-optimized or encoded probe with measured preparation, survival, accepted-information rate, and wall-clock resource accounting \\
\bottomrule
\end{longtblr}
\normalsize

\subsection{Where new astrophysics is most plausible}
The strongest qualitative opportunity is sub-Rayleigh orbital astrometry. A calibrated signed spatial-mode receiver could extend dynamical-mass measurements to closer binaries than conventional imaging can use efficiently, directly testing low-mass stellar and pre-main-sequence models. The calculation identifies a science case in which the quantum-informed basis changes the information carried by each detected photon; Eq.~\eqref{eq:pmsselect} shows the resulting leverage on an illustrative pre-main-sequence model decision.

Squeezed metrology and modal coronagraphy are more conditional. Squeezing already improves the relevant internal phase observable, yet the Earth-analog mass benchmark shows that additional squeezing is not the dominant route to new science unless the coupled readout fraction, zero point, stellar activity, and duty are improved together. Modal coronagraphy can be astrophysically important near the inner working angle, but only after finite-star leakage, time-dependent wavefront errors, throughput, crosstalk, and backgrounds are placed on a common target-level exposure calculation.

\subsection{Where the present architecture is the bottleneck}
Quantum heterodyne reception is attractive when coherent spectroscopy or distributed phase architecture is required, but its gain over classical heterodyne should not be confused with a photon-sensitivity advantage over direct detection. Entanglement-assisted reception remains a compelling long-baseline concept, yet the present model lacks the phase diversity and completed operations required to infer source diameter or brightness structure. The nuller and NOON cases show the same lesson in different forms: reducing a subdominant noise term or improving an ideal phase response does not create new astrophysics when another physical resource controls the final observable.

\subsection{Decisive next experiments}
\begin{enumerate}[leftmargin=*,itemsep=2pt]
\item \textbf{Earth-analog astrometry:} demonstrate a calibrated picometer differential-delay measurement with classical and squeezed schedules, and report the recovered angular signal after zero-point, duty, and activity terms are included.
\item \textbf{Young binaries:} recover a rotating two-source orbit with a broadband signed modal receiver under measured transfer drift and a fully independent direct-image likelihood.
\item \textbf{Exoplanet imaging:} measure modal-coronagraph throughput and finite-star leakage simultaneously under wavefront and exozodiacal backgrounds, then propagate the result to a planet spectrum or localization posterior.
\item \textbf{Coherent reception:} compare quantum heterodyne, classical heterodyne, and direct detection on the same weak thermal source and resource denominator.
\item \textbf{Nonlocal interferometry:} implement a phase-diverse visibility likelihood on a compact source and close pair delivery, memory, false-event, retry, and phase-reference resources per completed information-bearing operation.
\end{enumerate}

The Earth-twin fixed-time gain $G_{\rm fixed}=1.010$ and the internally calibrated SPADE value $G_{\rm matched}=1.77$ are not equivalent end-to-end quantities. The former includes a fuller duty and nuisance propagation; the latter remains a representative matched-model comparison. Detailed comparator, convergence, calibration, and interval definitions are retained in Appendix~\ref{app:spade} rather than used as the organizing narrative of the paper.

\section{Conclusions}
Quantum technologies do not improve optical astronomy in a generic way; they improve particular observables. The scientific question is whether that observable still controls the measurement after optical loss, calibration, observing duty, source structure, and astrophysical backgrounds are included.

For Earth-analog astrometry, squeezed internal metrology demonstrably improves the phase readout, but the mass measurement is already limited mainly by zero point, cadence geometry, stellar photocenter motion, and external orbital information. The modeled fixed-time gain is therefore only \EarthFixedGain\, and the change in the 10\%-mass characterization fraction is unresolved. Under this allocation, squeezing is an incremental instrument improvement rather than a new astrophysical capability.

Signed spatial-mode detection is the most promising positive case because it changes the information-bearing measurement basis. The representative young-binary calculation carries preserved sub-Rayleigh separation information through a two-dimensional orbit to a dynamical-mass RMSE gain \SpadeGain. In the explicit equal-prior illustration, 15\%, 10\%, and 7\% mass precision gives 73\%, 82\%, and 90\% correct selection between 0.10 and \(0.12M_\odot\) pre-main-sequence predictions. The exact \(1.77\times\) factor remains provisional pending the four closure tests defined in Sec.~\ref{sec:spade}; it should be treated as a representative benchmark rather than a universal receiver constant.

Modal coronagraphy could improve direct exoplanet imaging near the inner working angle, but only where planet throughput outweighs finite-star, instrumental, and modal leakage relative to a realistic conventional frontier. Quantum-enhanced heterodyne reception improves classical coherent detection yet does not overcome direct detection in the weak optical mode studied here; its value is more likely to lie in spectral resolution, frequency stability, or distributed architecture. Entanglement-assisted reception would open a qualitatively different long-baseline architecture, but the present stress case lacks the phase-diverse likelihood, useful correlated flux, and completed-operation yield required for source-structure science. The nuller and prepared-NOON controls show why a component-level quantum gain can be astrophysically irrelevant when another term dominates.

The practical design principle is therefore astronomy first: identify the source parameter and observable, establish the strongest classical measurement, and add quantum hardware only when the modified channel remains a dominant route to the desired astrophysical information. The paper's main outcome is an opportunity map for that decision---close-binary spatial-mode astrometry as the strongest current positive case, squeezed astrometry and modal coronagraphy as conditional opportunities, and coherent, nonlocal, nulling, and prepared-probe architectures as cases whose scientific value depends on closing specific physical bottlenecks.

%\section*{Data and software availability}
%The accompanying reproducibility archive contains this complete manuscript, the bibliography, publication and source figures, figure- and table-generation scripts, analytical and simulation code, configurations, random seeds, run manifests, and the retained raw and processed outputs used by the reported calculations. Deterministic analytical checks, receiver-specific convergence calculations, all publication figures, all tables, and the TeX manuscript can be regenerated from the archive. The full nonlinear Monte Carlo campaigns are provided with their code, inputs, seeds, retained outputs, and explicit execution commands; because they are computationally intensive, they are not part of the default reconstruction command. No persistent public repository identifier or DOI is claimed.

\begin{acknowledgments}
The work described here was carried out at the Jet Propulsion Laboratory, California Institute of Technology, Pasadena, California, under a contract with the National Aeronautics and Space Administration.  \copyright\ 2026 California Institute of Technology. Government sponsorship acknowledged.
\end{acknowledgments}

\appendix

\section{Comparator comparison and residual uncertainties}
\label{app:comparators}
The consistency check separates the immediate receiver comparator from the strongest modeled system-level alternative. Tables~\ref{tab:comparatorcomparison-a} and \ref{tab:comparatorcomparison-b} record, respectively, the common denominator and the architecture-specific charges, numerical optimization, and residual uncertainty.

\subsection{Likelihood propagation and resource accounting}
The following common notation is used only to define how a receiver change is propagated; the science-facing results are stated in the units of each science observable.

For receiver outcomes \(y\), parameter vector \(\boldsymbol\vartheta\), and likelihood \(p(y|\boldsymbol\vartheta)\), the classical Fisher matrix is
\begin{equation}
F_{ab}=\sum_y \frac{1}{p(y|\boldsymbol\vartheta)}
\frac{\partial p}{\partial\vartheta_a}
\frac{\partial p}{\partial\vartheta_b}.
\label{eq:fisher}
\end{equation}
For a balanced two-output combiner with complex source/instrument visibility \(V e^{i\phi}\),
\begin{equation}
p_\pm=\frac{1\pm V\cos\phi}{2},\qquad
F_\phi=N\frac{V^2\sin^2\phi}{1-V^2\cos^2\phi}.
\label{eq:phasefi}
\end{equation}
The familiar \(F_\phi=NV^2\) therefore applies at the tracked quadrature point \(\phi=\pi/2\).  With Gaussian prior precision \(P\), the posterior covariance is
\begin{equation}
C=(F_{\rm data}+P)^{-1};
\label{eq:postcov}
\end{equation}
nonlinear benchmarks instead use their full Poisson or Gaussian-process likelihood and empirical recovery.

For a finite covariance intervention \(q_0\rightarrow q_1\), we define the diagnostic
\begin{equation}
\Efin=
\frac{1-\sigma_a^2(q_1)/\sigma_a^2(q_0)}{1-q_1/q_0},
\label{eq:efin}
\end{equation}
where both target variances are obtained by rerunning the complete nuisance model.  It is an intervention-specific sensitivity, not a unique decomposition of a correlated posterior.  Its local differential analogue follows from \(C=J^{-1}\):
\begin{equation}
\frac{\partial C}{\partial q}=-C\frac{\partial J}{\partial q}C,
\qquad
E_q=-\frac{q}{\sigma_a^2}\,
\bm g^{\mathsf T}C\frac{\partial J}{\partial q}C\bm g,
\label{eq:elasticity}
\end{equation}
with \(\sigma_a^2=\bm g^{\mathsf T}C\bm g\).  The finite and local quantities agree only for sufficiently small, path-independent changes.

\subsection{Resource denominators and comparator logic}
The resource denominator is architecture dependent.  Active metrology is compared at fixed carrier, accepted bandwidth, detector chain, and elapsed time; squeezed photons, pump and lock burden, optical loss, and duty are charged separately.  Measurement-basis receivers use the same source photons, aperture, spectrum, background, and observing time while charging receiver throughput, calibration, and detector differences.  Passive nonlocal reception uses the same collected sky photons as the classical link but adds Bell pairs, memories, gates, switching, phase reference, false events, and retries.  Equal spacecraft mass, wall power, cooling, reliability, lifecycle cost, and schedule are not claimed where no physical implementation supplies credible scaling laws.

Science leverage is therefore distinct from cost effectiveness.  If \(q_j\) is an engineering parameter and \(C_j\) a common cost coordinate,
\begin{equation}
\frac{\partial\ln\sigma_{\rm sci}^2}{\partial C_j}
=
\frac{\partial\ln\sigma_{\rm sci}^2}{\partial\ln q_j}
\frac{\partial\ln q_j}{\partial C_j}.
\label{eq:costderivative}
\end{equation}
This work calculates the first factor and finite intervention gains.  The second requires a flight-architecture model.  Comparisons below consequently answer which physical intervention has greater science leverage under the declared change, not which yields the greatest return per dollar or kilogram.

The provenance of the load-bearing numerical inputs is reconstructed in Appendix~\ref{app:provenance}.  Literature-derived physical models, demonstrated mechanisms, design targets, illustrative allocations, emulator ranges, and derived requirements are labeled separately so that a favorable point estimate is not mistaken for a measured flight allocation.

\small
\setlength{\tabcolsep}{3.0pt}\renewcommand{\arraystretch}{1.08}
\begin{longtblr}[caption={Comparator comparison, part I: immediate comparator and common resource basis.  ``System alternative'' names the strongest modeled non-quantum intervention beyond the immediate receiver comparator.},label={tab:comparatorcomparison-a}]{width=\textwidth,colspec={X[13,l] X[18,l] X[27,l] X[36,l]},rowhead=1}
\toprule
Approach & Immediate comparator & System alternative & Quantities held fixed \\\midrule
Squeezed metrology & Unsqueezed homodyne & Mission duration, baseline, cadence, zero point, and activity mitigation & Carrier, accepted band, detector chain, candidate schedule, source likelihood, and elapsed time \\
SPADE & Pixel-integrated direct imaging & Continuous or unbinned point-spread-function likelihood & Source photons, aperture, spectrum, backgrounds, calibration draws, and orbit priors \\
Modal coronagraphy & Conventional coronagraph & Better wavefront control and background rejection & Pupil scale, wavelength, separation, contrast, background convention, and exposure \\
Quantum heterodyne & Classical heterodyne & Direct photon detection & Thermal source occupation, background, optical band, detector model, and integration time \\
Nonlocal reception & Coherent relay & Heterodyne or HBT where the observable is compatible & Source, channels, apertures, observing time, background, and inference priors \\
Nuller sensor & Classical controller & More aperture, lower exozodiacal background, and better disturbance rejection & Plant, scene, exposure, controller constraints, stability margins, and actuator limits \\
Canonical NOON & Coherent phase probe & Increased coherent-photon budget & Attempted photons, path, detector technology, and wall-clock duration \\
\bottomrule
\end{longtblr}

\begin{longtblr}[caption={Comparator comparison, part II: architecture-specific charges, optimization, and residual uncertainty.},label={tab:comparatorcomparison-b}]{width=\textwidth,colspec={X[11.5,l] X[26.5,l] X[28,l] X[32,l]},rowhead=1}
\toprule
Approach & Architecture-specific charges & Optimization and convergence & Residual uncertainty \\\midrule
Squeezed metrology & Optical loss, squeeze-angle lock, pump/control boundary, and valid-epoch duty & Same nonlinear fit under paired accepted-data and fixed-time schedules & Pump and wall-power architecture remain parametric; the joint design envelope places substantial probability near unit gain \\
SPADE & Mode sorter, lower detector efficiency, registration allocation, and calibration data & Fifteen grid/field tests, high-resolution paired refinement, and local unbinned-information benchmark & Strict tail-sensitive convergence is not met; the exact gain remains comparator and calibration dependent \\
Modal coronagraphy & Modal projection, planet throughput, finite-star leakage, and crosstalk & Analytical Poisson receiver comparison & Conventional leakage/throughput frontier and time-dependent wavefront control determine break-even \\
Quantum heterodyne & Squeezed ancillary field and local-oscillator/reference distribution & Analytical Gaussian receiver in the weak-occupation domain & Array-wide local-oscillator power and phase resources are not closed; direct detection dominates the selected source regime \\
Nonlocal reception & Bell pairs, memories, gates, conversion, switching, phase reference, and false events & Fresnel dominant-mode transport and terminal-diameter sweep & No physical encoded receiver is specified; pair supply, memory, and useful source coherence remain open \\
Nuller sensor & Squeezed-sensor loss, lock, and duty & Independent classical and squeezed controller optimization with nested refinement & Reduced-order plant and background model; used as a finite-ensemble negative control \\
Canonical NOON & Preparation, heralding, ancillas, all-photon survival, phase diffusion, and discarded trials & Orders 1--20 evaluated under one attempted-photon accounting & Does not classify optimized or encoded non-Gaussian probes \\
\bottomrule
\end{longtblr}
\normalsize

\section{Engineering assumptions and parameter provenance}
\label{app:provenance}
Tables~\ref{tab:provenance} and \ref{tab:provenanceb} summarize the provenance of the load-bearing inputs. The provenance vocabulary is controlled throughout the manuscript.  \emph{Measured} denotes a value obtained from the particular hardware or data analyzed here.  \emph{Experimentally demonstrated} denotes a capability shown in a relevant experiment but not necessarily in the present architecture.  \emph{Literature-derived physical model} denotes an adopted primary-source relation or range.  A \emph{design target} is an intended future allocation, an \emph{illustrative assumption} is a selected benchmark without a maturity claim, an \emph{emulator range} is used only for sensitivity exploration, and a \emph{derived quantity or requirement} follows mathematically from declared inputs.  A demonstrated physical mechanism does not make every downstream engineering allocation measured.  The complete machine-readable consistency check, including demonstrated context, manuscript context, extrapolation, affected output, sensitivity, and evidence needed to replace an assumption, is supplied in the machine-readable parameter-provenance table in the accompanying reproducibility archive.

\small
\setlength{\tabcolsep}{3pt}\renewcommand{\arraystretch}{1.08}
\begin{longtblr}[caption={Consolidated provenance of load-bearing Earth-twin and SPADE inputs.  ``Sensitivity'' is the qualitative effect on the headline conclusion within the tested model.},label={tab:provenance}]{width=\textwidth,colspec={X[7,l] X[14,l] X[14,l] X[13,l] X[15,l] X[28,l] X[10,l]},rowhead=1}
\toprule
Case & Parameter & Nominal value & Tested range & Category & Source/derivation & Sensitivity \\\midrule
Earth & Input squeezing & 6 dB & 3--10 dB & Design target & Operational mechanism demonstrated in large interferometers~\cite{Tse2019,Acernese2019,Ganapathy2023}; space allocation extrapolated & Moderate \\
Earth & Total optical efficiency & 0.80 & 0.68--0.92 & Design target & Product of injection, propagation, mode matching, readout, and detector efficiencies & High \\
Earth & Squeeze-angle jitter & 0.02 rad RMS & 0.004--0.055 & Design target & Closed-loop receiver-stability allocation & Moderate \\
Earth & Quantum-coupled readout & 25 pm within 35 pm & coupling 0.25--0.82 & Illustrative assumption & Noise-component decomposition in Eq.~\eqref{eq:beta} & High \\
Earth & Zero-point sharing & 9\% of variance & 0--30\% & Illustrative assumption & Fraction of zero point assumed to share the enhanced optical path & High \\
Earth & Valid-operation duty & 0.980 / 0.932 & squeezed 0.86--0.985 & Design target & Acquisition, lock, calibration, and recovery accounting & High \\
Earth & Photocenter process & 0.05 $\mu$as, 25 d & scale 0.5--2.2 & Literature-derived physical model & Solar/starspot astrometric-jitter models~\cite{Makarov2009Starspot,Deagan2026SolarJitter} & High \\
Earth & RV precursor & $\sigma_K=0.10$ m s$^{-1}$ & 0.03--0.28 & Illustrative assumption & 120-visit quasi-periodic campaign calculation & High \\
Earth & Joint design envelope & declared triangular ranges & see App.~\ref{app:earthsens} & Emulator range & 60,000 finite-intervention samples & Diagnostic \\
SPADE & Optical model & Gaussian, $0.437\lambda/D$ & local Airy consistency check & Illustrative assumption & End-to-end Gaussian-PSF benchmark & High \\
SPADE & Detector efficiencies & 0.90 direct / 0.85 modal & fixed & Design target & Receiver photon-rate accounting & Moderate \\
SPADE & Registration allocation & 25\% detected photons & 10--40\% & Illustrative assumption & Signed-direction and sorter-origin channel & High \\
SPADE & Nominal crosstalk & $2\times10^{-4}$ & 0--0.8\% & Design target & Nearest-neighbor transfer matrix & High \\
SPADE & Transfer drift & discrete origin, angle, gain, and crosstalk cases & 45-d correlation & Emulator range & Held-out calibration stress tests & High \\
SPADE & Target radiometry & AB 16, 3200 K; $1.23\times10^5$ photons/epoch & eight-system population & Illustrative assumption & Eq.~\eqref{eq:spaderad} and target-population file & Moderate \\
\bottomrule
\end{longtblr}

\begin{longtblr}[caption={Provenance of receiver-level and feasibility benchmarks.},label={tab:provenanceb}]{width=\textwidth,colspec={X[12,l] X[16,l] X[16,l] X[16,l] X[22,l] X[10,l]},rowhead=1}
\toprule
Case & Parameter & Nominal value & Category & Source/derivation & Output \\\midrule
Modal coronagraph & Planet throughput & 0.500648 & Derived quantity or requirement & Modal-overlap calculation & Receiver speed \\
Modal coronagraph & Finite-star leakage & $3.47\times10^{-5}$ & Derived quantity or requirement & Finite-star projection & Noise balance \\
Modal coronagraph & Modal crosstalk & $10^{-4}$ & Illustrative assumption & Conditional frontier & Speed ratio \\
Quantum heterodyne & Image-band variance & 0.447 & Experimentally demonstrated & Laboratory receiver mechanism~\cite{Gould2024Heterodyne,Anai2024Heterodyne} & Receiver FI \\
Network & Per-operation efficiencies & 0.55 / 0.55 / 0.90 / 0.75 & Illustrative assumption & Unencoded operation model & Completed yield \\
Network & Phase-diverse amplitude schedule & not specified & Unresolved & No phase-diverse schedule in declared model & No amplitude crossover \\
Nuller & Plant/controller & 30-nm disturbance; 5-ms delay & Illustrative assumption & Reduced-order finite-ensemble model & Instrument proxy \\
NOON & Preparation, duty, efficiency & 0.05/0.70/0.80 & Illustrative assumption & Attempted-photon-rate accounting & Receiver gain \\
\bottomrule
\end{longtblr}
\normalsize

\section{Earth-twin sensitivity-emulator definition}
\label{app:earthsens}
The 60,000-sample design-envelope calculation is a finite-intervention emulator calibrated to the nonlinear accepted-data result; it is not a second nonlinear posterior.  Triangular distributions use minimum/mode/maximum values: squeezing 3/6/10 dB, efficiency 0.68/0.80/0.92, angle jitter 0.004/0.02/0.055 rad, readout coupling 0.25/0.510/0.82, zero-point coupling 0/0.09/0.30, squeezed duty 0.86/0.932/0.985, readout scale 0.80/1/1.25, zero-point scale 0.50/1/1.60, photocenter scale 0.50/1/2.20, and thermal scale 0.50/1/1.60.  The nominal accepted-data finite sensitivity \(\Efin=0.10382\) is rescaled by the ratio of quantum-coupled readout/zero-point covariance to the total finite-intervention burden.  The fixed-time gain includes the calibrated nonlinear geometry factor and \(\sqrt{d_q/d_{\rm cl}}\).  Seed 2026082036 and all 60,000 samples are included in the reproducibility archive.  The model is useful for identifying joint dependencies and probability mass within declared ranges, but any flight prediction requires distributions derived from hardware and astrophysical measurements.

\section{SPADE implementation and validation details}
\label{app:spade}
The source radiometry is integrated per channel using a 3200-K spectral shape normalized to total AB magnitude 16.  The weak color law gives component flux ratios 0.693, 0.700, and 0.707; holding the ratio fixed changes no qualitative conclusion.  Registration pixels are integrated over the Gaussian PSF and contain an explicit out-of-field probability.  The matched calibration state includes static crosstalk \(2\times10^{-4}\), residual leakage \(10^{-4}\), modal gain RMS \(2\times10^{-3}\), and finite background.  Five-fold interval scaling uses held-out folds within the same simulator family.

The direct-grid consistency check covers \(5\times5\), \(7\times7\), \(9\times9\), \(11\times11\), \(13\times13\), \(17\times17\), and \(21\times21\) configurations; the comparator table uses the common \(13,17,21\) high-resolution, \(\pm5\sigma_{\rm PSF}\) data.  The available sample accounting is reproduced in Table~\ref{tab:spaderecord}. Point estimates and intervals are paired only when a common-realization file exists; the 160-trial 21-grid summary has no 160-pair common-run file and is therefore not assigned a confidence interval. This asymmetry is why no exact converged factor is claimed.

\subsection*{Uncertainty-interval and coverage definitions}
Unless stated otherwise, a reported bootstrap interval is the two-sided 95\% percentile interval.  Resampling never mixes receiver labels: a paired draw selects one stored physical-system or trial index and retains all receiver-specific outcomes attached to that index.  Catastrophic recoveries and other finite fitted outcomes remain in the estimand.

\paragraph{Earth-twin mass-precision gains.}
For accepted-data and fixed-time inference, the estimand is the ratio of the sample medians of per-realization fractional mass uncertainty, classical divided by squeezed.  The resampling unit is one paired physical-system realization indexed by the common \((\text{seed},\text{trial})\) identifier; the same 160 indices are used for both architectures.  The preserved provenance record specifies 20,000 paired percentile resamples and seed 4504, with all 160 successful fitted pairs retained and each realization's cadence, nuisance draw, and noise outcome kept together.  These definitions apply to the intervals in the abstract, Fig.~\ref{fig:earthGain}(a), the Earth-analog section, and the associated text.  The original resampled-index arrays and the exact original routine ordering were not archived, so the manuscript does not claim draw-by-draw replay of those two intervals beyond the stored endpoints, algorithm, resample count, and seed.

\paragraph{Earth-twin requirement curves.}
At each mass-precision requirement \(r\), the paired estimand is
\(N^{-1}\sum_i[\mathbf 1(m_{q,i}\le r)-\mathbf 1(m_{{\rm cl},i}\le r)]\).
The 160 paired physical-system indices are sampled with replacement, the same index is applied to both architectures, and the mean indicator difference is recomputed.  Figure~\ref{fig:earthGain}(d) uses 2,500 percentile resamples from one generator stream with seed 2026081930 while the requirements are evaluated in ascending order; all 160 successful pairs are retained.  The architecture-specific bands in Fig.~\ref{fig:earthGain}(c) are not bootstrap intervals: they are two-sided 95\% Wilson intervals with \(z=1.9599639845\) for the 160 Bernoulli requirement outcomes in each architecture.  Auxiliary error bars in Fig.~\ref{fig:earthdesign}(a) are retained from stored paired-ensemble summaries; the archived calculation records do not preserve a complete seed-and-index specification for every auxiliary intervention, so those bars are not claimed to have an independently replayable resampling sequence.

\paragraph{SPADE matched-comparator intervals.}
For every row in Tables~\ref{tab:spadeconv} and \ref{tab:spaderecord}, the RMSE-gain estimand is \(\operatorname{RMSE}(e_{\rm DI})/\operatorname{RMSE}(e_{\rm SP})\), and the MedAE-gain estimand replaces RMSE by \(\operatorname{median}|e|\).  Each interval uses 10,000 paired percentile resamples of common trial indices, with the same resampled indices applied to direct imaging and SPADE.  The RMSE/MedAE seeds are, respectively: \(15195/15196\) for \(13\times13, N=160\); \(19075/19076\), \(19115/19116\), and \(2035/2036\) for \(17\times17, N=40,80,160\); and \(23051/23052\), \(23067/23068\), and \(23083/23084\) for \(21\times21, N=16,32,48\).  All common fitted pairs, including catastrophic recoveries, are retained.  The implementation excludes only a nonfinite MedAE-ratio bootstrap draw if a resampled SPADE MedAE is exactly zero.  The canonical Fig.~\ref{fig:spadeinf} and Eq.~\eqref{eq:spadeGains} RMSE interval therefore use 10,000 common-index resamples with seed 2035.

\paragraph{SPADE crosstalk, calibration, coverage, and population intervals.}
The five static-crosstalk intervals in Fig.~\ref{fig:spadecal}(a) use the RMSE-gain estimand, 48 complete paired photon-realization trials per tested crosstalk point, 3,000 paired percentile resamples, and seed 55 at each point; crosstalk levels themselves are not resampled or interpolated.  The separated calibration and combined-drift bars in Fig.~\ref{fig:spadecal}(b) likewise resample 48 complete paired trial indices 3,000 times and recompute the RMSE-gain ratio.  The stored data and routine establish the estimand, unit, pairing, interval type, and replicate count, but the exact seed mapping that generated the retained bar endpoints is not recoverable from the archived calculation records; an independently reproducible draw sequence is therefore unavailable for those bars.  The combined-drift 68\% coverage statement is a Wilson interval for 20 covered outcomes among 48 trials, not a bootstrap interval.  The population interval in Fig.~\ref{fig:spadecal}(d) is a cluster percentile bootstrap of the RMSE-gain ratio: eight physical systems are sampled with replacement, all four photon realizations belonging to each selected system are retained, 10,000 cluster draws are used, and the seed is 2026082038.  No individual photon realization is resampled independently of its physical-system cluster.

\paragraph{Nuller percentile interval.}
The nuller estimand is the sample 95th percentile of \(10^6(\sigma_{\rm cl}-\sigma_q)/\sigma_{\rm cl}\) across the 180 held-out plant/background realizations.  Bootstrap draws resample those realizations with replacement and recompute the 95th percentile; all 180 finite realizations are retained.  The supplied output preserves the interval endpoints but not the original bootstrap replicate count, random seed, or resampled-index sequence.  Figure~\ref{fig:nuller} and its accompanying text therefore report the stored percentile interval while explicitly noting that its draw sequence is not independently reproducible from the archived calculation records.

% Inlined table definition.
%\small
\setlength{\tabcolsep}{2.8pt}\renewcommand{\arraystretch}{1.10}
\begin{longtblr}[caption={SPADE/direct-imaging paired statistical record. RMSE and MedAE are fractional dynamical-mass errors. The gain point estimate is the ratio of the unrounded direct and SPADE statistics; intervals are paired-bootstrap 95\% percentile intervals.},label={tab:spaderecord}]{width=\textwidth,colspec={X[28,l] X[6,c] X[10,r] X[10,r] X[10,r] X[10,r] X[20,l]},rowhead=1}
\toprule
\small
Comparator & {$N_{\rm pair}$} & {RMSE$_{\rm DI}$} & {RMSE$_{\rm SP}$} & {MedAE$_{\rm DI}$} & {MedAE$_{\rm SP}$} & Gain [95\% interval]\\\midrule
$13\times13$, $\pm$5 $\sigma_{\rm PSF}$ & 160 & 0.3537 & 0.1810 & 0.1947 & 0.1251 & 1.95 [1.58--2.39]\\
$17\times17$, $\pm$5 $\sigma_{\rm PSF}$, first 40 common pairs & 40 & 0.2906 & 0.1953 & 0.1962 & 0.1549 & 1.49 [1.16--1.83]\\
$17\times17$, $\pm$5 $\sigma_{\rm PSF}$, first 80 common pairs & 80 & 0.2819 & 0.1874 & 0.1880 & 0.1304 & 1.50 [1.21--1.85]\\
$17\times17$, $\pm$5 $\sigma_{\rm PSF}$, first 160 common pairs & 160 & 0.3212 & 0.1810 & 0.1786 & 0.1251 & 1.77 [1.47--2.15]\\
$21\times21$, $\pm$5 $\sigma_{\rm PSF}$, first 16 common pairs & 16 & 0.3319 & 0.1951 & 0.2742 & 0.1574 & 1.70 [1.05--2.84]\\
$21\times21$, $\pm$5 $\sigma_{\rm PSF}$, first 32 common pairs & 32 & 0.2946 & 0.2060 & 0.2052 & 0.1574 & 1.43 [1.00--2.00]\\
$21\times21$, $\pm$5 $\sigma_{\rm PSF}$, first 48 common pairs & 48 & 0.3650 & 0.1932 & 0.2245 & 0.1498 & 1.89 [1.35--2.57]\\
\bottomrule
\end{longtblr}
%\normalsize
%\footnotesize
\small
\setlength{\tabcolsep}{2.4pt}\renewcommand{\arraystretch}{1.12}
\begin{longtblr}[caption={Receiver-specific SPADE/direct-imaging sequential convergence record.  Relative changes in RMSE, MedAE, and RMSE gain pass only when strictly below 5\%; absolute changes in catastrophic fraction and raw 68\% coverage pass only when strictly below 2 percentage points.  Each entry gives value followed by P or F.  All values are computed from the unrounded row statistics in Table~\ref{tab:spaderecord}.},label={tab:spadetail}]{width=\textwidth,colspec={X[18,l] X[19,c] X[19,c] X[10,c] X[18,c] X[18,c] X[6,c]},rowhead=1}
\toprule
Transition & {$\Delta$RMSE DI/SP (\%)} & {$\Delta$MedAE DI/SP (\%)} & {$\Delta G_{\rm RMSE}$ (\%)} & {$\Delta$cat. DI/SP (pp)} & {$\Delta$raw cov. DI/SP (pp)} & Overall\\\midrule
$17\times17$: $40\to80$ & 2.985 P / 4.056 P & 4.136 P / 15.861 F & 1.116 P & 0.000 P / 1.250 P & 2.500 F / 3.750 F & Fail\\
$17\times17$: $80\to160$ & 13.915 F / 3.405 P & 5.0379 F / 4.023 P & 17.931 F & 0.625 P / 0.625 P & 3.125 F / 5.000 F & Fail\\
$21\times21$: $16\to32$ & 11.239 F / 5.602 F & 25.183 F / 0.000 P & 15.948 F & 0.000 P / 0.000 P & 3.125 F / 6.250 F & Fail\\
$21\times21$: $32\to48$ & 23.900 F / 6.208 F & 9.448 F / 4.886 P & 32.101 F & 8.333 F / 0.000 P & 7.292 F / 2.083 F & Fail\\
\bottomrule
\end{longtblr}
\normalsize

The orbit solves Kepler's equation and projects the eccentric orbit through inclination, node, and argument of periastron.  Total mass follows \(M_{\rm tot}=a_{\rm AU}^3/P_{\rm yr}^2\).  Directional sweeps vary flux ratio, orientation-prior width, registration fraction, origin error, and sorter angle.  The population interval resamples the eight physical systems as clusters and retains their four photon realizations together.

\section{Receiver-level map definitions}
\label{app:maps}
For modal coronagraphy, the reference accounting is planet contrast \(10^{-5}\), separation \(0.7\lambda/D\), stellar diameter \(0.014\lambda/D\), background \(10^{-6}\) of the stellar rate, modal planet throughput 0.5006, finite-star leakage \(3.47\times10^{-5}\), modal crosstalk \(10^{-4}\), conventional throughput 0.30, and conventional leakage \(10^{-4}\).  Equation~\eqref{eq:coro} assumes Poisson statistics and does not include a time-dependent wavefront controller.

The heterodyne map is restricted to \(\epsilon,n_b\le10^{-2}\), where Eq.~\eqref{eq:qhetdirect} is a weak-occupation approximation.  \(V_{\rm img}=0.447\) is a variance ratio, so \(G_{\rm FI}\) and \(G_\sigma\) differ by a square root.  Local-oscillator power, comb distribution, cryogenics, and array-wide phase transfer are outside the receiver map.  Enabling hardware should be evaluated separately from nonclassical-state advantage: integrated memories can improve storage efficiency and bandwidth~\cite{Meng2026}, while radiation tolerance can dominate the space utility of photon-counting hardware~\cite{Lenart2025}.

\section{Network likelihood and protocol consistency calculation}
\label{app:network}
The 300 channels span 0.995--1.005 $\mu$m uniformly in frequency. The Sirius-B total incident rate is $2.28\times10^4\,\mathrm{s^{-1}}$ per 0.5-m telescope. The event likelihood is Eq.~\eqref{eq:networkrates}. Equation~\eqref{eq:networkphasefi} is verified analytically and by finite differences for a phase parameter; Eq.~\eqref{eq:networkampfi} is the corresponding visibility-amplitude result and vanishes at quadrature. The declared model contains no phase-step configuration for visibility-amplitude inference. Consequently no diameter-information boundary, pair-rate requirement, or memory requirement derived from such a boundary is reported.

The Fresnel relay calculation itself remains a valid transport model~\cite{GoodmanFourier2005}: it uses the dominant singular mode between circular pupils with 0.20-$\mu$rad pointing RMS and 30-nm wavefront RMS. Heterodyne efficiency is 0.45; HBT efficiency is 0.80 with 100-ps timing. These comparators measure different statistics and are not collapsed into one universal photon efficiency. Equation~\eqref{eq:yunc} counts one already aggregated write, read, detection, and gate operation for each of 23 address bits. A physical encoded receiver must additionally define its phase schedule, code, logical gate sequence, retry/heralding policy, unheralded-error model, false heralds, source multipairs, memory fluorescence, conversion and switching loss, phase-reference distribution, and cooling/power architecture.

\section{Nuller and prepared-probe validation}
\label{app:controls}
The nuller disturbance spectrum is a low-frequency rolloff plus a 5-Hz structural line normalized to 30-nm RMS.  The white sensor spectrum integrates to 0.45 nm over \(10^{-3}\)--200 Hz; actuator noise has a 30-Hz rolloff and integrates to 0.70 nm.  Four nested refinements follow the initial controller grid, rejecting candidates that violate phase margin, gain margin, or command RMS.  The 180 held-out realizations vary plant delay/pole, disturbance amplitude, structural-line frequency, and background factors with the selected controllers frozen.  The reported interval resamples realizations and recomputes the sample 95th percentile.  The stored output does not retain the bootstrap seed, replicate count, or index sequence, so exact resampling replay is unavailable even though the 180 input realizations and interval endpoints are preserved.

The NOON calculation uses attempted photons as the denominator.  Preparation failures, lost photons, detector inefficiency, and discarded trials therefore count against elapsed time.  \(\kappa\) converts ancillary and attempt overhead to the coherent-probe resource convention.  Accepted-photon normalization would answer a different, postselected question and is not used for the conclusion.

\bibliographystyle{apsrev4-2}
%\bibliography{quantum_interferometry}

%apsrev4-2.bst 2019-01-14 (MD) hand-edited version of apsrev4-1.bst
%Control: key (0)
%Control: author (72) initials jnrlst
%Control: editor formatted (1) identically to author
%Control: production of article title (-1) disabled
%Control: page (0) single
%Control: year (1) truncated
%Control: production of eprint (0) enabled
%

\end{document}